\documentclass[twocolumn,superscriptaddress,floatfix,amsmath,amssymb,aps,prb]{revtex4-2}
\usepackage[utf8]{inputenc}
\usepackage{chemformula}
\usepackage{amsmath}
\usepackage{gensymb}
\usepackage{mathrsfs}
\usepackage{physics}
\usepackage{multirow}
\usepackage{amssymb}
\usepackage{graphicx}
\usepackage{siunitx}
\usepackage{lipsum}
\usepackage[normalem]{ulem}
\usepackage[colorlinks=true, allcolors=blue]{hyperref}
\usepackage{float}
\newcommand{\mrm}[1]{\mathrm{#1}}

\newcommand{\Tc}{T_{\mrm{c}}}
\newcommand{\Tcon}{\Tc^{\mrm{onset}}}
\newcommand{\Tcmid}{\Tc^{\mrm{mid}}}

\newcommand{\bk}{\mathbf{k}}
\newcommand{\ra}{\rangle}
\newcommand{\la}{\langle}

\newcommand{\mcolc}[1]{\multicolumn{1}{c}{#1}}
\newcommand{\cdash}{\mcolc{-}}

\newcommand{\Hire}{\cite{Hire2022}}
\newcommand{\D}{\mrm{d}}

\definecolor{mag}{RGB}{255,0,255}

\usepackage{xcolor}
\usepackage{pagecolor}
\usepackage{adjustbox}

\begin{document}
\title{Diboride compounds doped with transition metals\textemdash a route to superconductivity 
through structure stabilization as well as defects}
\date{\today}

\begin{abstract} 
Recent investigations into MoB$_{2}$ have unveiled a direct connection between a pressure-induced structural transition to a P$6/mmm$ space group structure and the emergence of superconductivity, producing critical temperatures up to 32 K at 100 GPa. 
This pressure-induced superconducting state underscores the potential of doped MoB$_{2}$ as a possible candidate for metastable superconductivity at ambient pressure. 
In this work, we demonstrate that doping by Zr, Hf, or Ta stabilizes the P$6/mmm$ structure at ambient pressure and results in the realization of a superconducting state with critical temperatures ranging from 2.4 up to 8.5 K depending on the specific doping. 
We estimate the electron-phonon coupling $\lambda$ and the density of states based on resistivity and specific heat data, finding that $\lambda$ ranges from 0.4 - 0.6 for these compounds.
Finally, to investigate the role of possible metastable defect structures on the critical temperature, we analyze MoB$_{2}$, MoB$_{2.5}$, and Nb/Zr-doped MoB$_{2}$ using rapid cooling techniques. 
Notably, splat-quenching produces samples with higher critical temperatures and even retains superconductivity in MoB$_{2}$ at ambient pressure, achieving a critical temperature of 4.5 K.\\

\end{abstract}
\author{P. M. Dee}
\thanks{These two authors contributed equally to this work.}
\affiliation{Department of Physics, University of Florida, Gainesville, Florida 32611, USA}
\affiliation{Department of Materials Science and  Engineering, University of Florida, Gainesville, Florida 32611, USA}
\author{J. S. Kim}
\thanks{These two authors contributed equally to this work.}
\affiliation{Department of Physics, University of Florida, Gainesville, Florida 32611, USA}
\author{A. C. Hire}
\affiliation{Department of Materials Science and  Engineering, University of Florida, Gainesville, Florida 32611, USA}
\affiliation{Quantum Theory Project, University of Florida, Gainesville, Florida 32611, USA}
\author{J. Lim}
\affiliation{Department of Physics, University of Florida, Gainesville, Florida 32611, USA}
\author{L.\ Fanfarillo}
\affiliation{Department of Physics, University of Florida, Gainesville, Florida 32611, USA}
\affiliation{Scuola Internazionale Superiore di Studi Avanzati (SISSA), Via Bonomea 265, 34136 Trieste, Italy}
\author{S. Sinha}
\affiliation{Department of Physics, University of Florida, Gainesville, Florida 32611, USA}
\author{J. J. Hamlin}
\affiliation{Department of Physics, University of Florida, Gainesville, Florida 32611, USA}
\author{R. G. Hennig}
\affiliation{Department of Materials Science and  Engineering, University of Florida, Gainesville, Florida 32611, USA}
\affiliation{Quantum Theory Project, University of Florida, Gainesville, Florida 32611, USA}
\author{P. J. Hirschfeld}
\affiliation{Department of Physics, University of Florida, Gainesville, Florida 32611, USA}
\author{G. R. Stewart}
\affiliation{Department of Physics, University of Florida, Gainesville, Florida 32611, USA}

\maketitle

\section{Introduction} 

The 2001 discovery of high-temperature superconductivity in MgB$_{2}$~\cite{Nagamatsu2001} reignited what had, until that point, been a latent interest in the superconducting properties of diborides.  
The resulting wave of new investigations explored alternatives to Mg using transition metals (TMs). Despite a broad sampling of TM's, the pursuit failed to unearth a worthy competitor. Similar to the findings of much earlier work~\cite{Tyan1969,Cooper1970,Leyarovska1979}, many of these TM-diborides were not superconductors or had critical temperatures ($\Tc$) below 10 K. As it stands, MgB$_{2}$ still retains the highest measured $\Tc$ of any diboride at 39 K. However, a recent study of MoB$_{2}$ has come surprisingly close to this title, with $\Tc=32$ K at high pressure~\cite{Pei2023MoB2}.

At ambient pressure, MoB$_{2}$ has no apparent superconductivity down to 1.8 K, and unlike MgB$_{2}$, every other boron layer is buckled (``puckered''), leading to the R$\bar{3}m$ [166] space group structure instead of the P6/$mmm$ [191] phase observed in MgB$_{2}$. However, above 20 GPa of applied pressure, a finite $\Tc$ emerges, ascending sharply at a rate of 0.7 K/GPa with increasing pressure until the system undergoes a structural phase transition to the P6$/mmm$ phase near 70 GPa. Thereafter, the $\D\Tc/\D P$ rate drops to 0.1 K/GPa,  with $\Tc$ eventually reaching 32 K near 100 GPa~\cite{Pei2023MoB2}. This hitherto unseen behavior in TM diborides raises questions about superconductivity in MoB$_{2}$ and whether it can be manipulated by pressure-induced metastability or partial substitutions with other TMs.
%

Early work by Cooper \textit{et al.}~\cite{Cooper1970} investigated the possibility of superconductivity in MoB$_{2}$ and several intermetallic boride compounds containing elements in the series Y, Zr, Nb, and Mo, mostly with boron concentrations above 2 compared with stoichiometric diborides. This exploration was partially motivated by the notion of an optimal electron/atom (e/a) ratio for superconductivity in these compounds. They claimed to find a correlation between the maximum observed $\Tc$'s and an e/a of 5 - 7. However, they did not observe superconductivity in either stoichiometric NbB$_{2}$ or MoB$_{2}$, even when the latter was synthesized using splat-quenching techniques. Only in the presence of excess boron\textemdash nominally reported as NbB$_{2.5}$ and MoB$_{2.5}$\textemdash did they find superconductivity, measuring the onset of $\Tc$ to be 6.4 K and 8.1 K, respectively. They further explored various alloyed diboride compounds by partially substituting Mo with another metal $M$ with nominal compositions given by Mo$_{2-x}M_{x}$B$_{5}$, finding $\Tc$'s ranging from 4.5 K to 11.2 K (the latter corresponding to Mo$_{1.69}$Zr$_{0.31}$B$_{5}$).       

The role of TM substitution in stabilizing the AlB$_{2}$ type structure in MoB$_{2+y}$ was further established by Muzzy \textit{et al.}~\cite{Muzzy2002}. Using Zr-substitution near 4\%, they created MoB$_{2}$ alloys in a metastable AlB$_{2}$ structure, obtaining compositions of the form (Zr$_{0.04}$Mo$_{0.96}$)$_{x}$B$_{2}$. By increasing the ratio of excess boron, they found that the samples harbored more metal vacancies, and the stoichiometric diboride phase showed evidence of $c$-axis stacking defects. Detecting the superconducting $\Tc$ from magnetometer measurements of the magnetization in an applied field, they found that $\Tc$ increases from about 5.9 K for $x=1.0$ to 8.2 K for $x=0.85$. In addition to having the lowest $\Tc$, the stoichiometric compound Zr$_{0.04}$Mo$_{0.96}$B$_{2}$ had the broadest transition, which the authors attributed to stacking defects and associated strains. 

Recent studies have explored the effects of Sc~\cite{Yang2022} and Nb~\cite{Hire2022,Lim2023suppression} substitutions in compounds of the form ($M_{y}$Mo$_{1-y})_{x}$B$_{2}$. Like Zr, Sc, and Nb possess fewer $d$-electrons than Mo, which results in a weaker overlap between the $d_{z^{2}}$ orbitals across TM-layers and between the TM-layer and adjacent boron sublattice. Consequently, the bond strength between boron atoms diminishes, expanding the intralayer boron atom separation and causing the alternating puckered boron-layers present in the rhombohedral phase of MoB$_{2}$, to flatten~\cite{Burdett1986}. In the study by Yang et al.~\cite{Yang2022}, the metal-deficient composition (Sc$_{0.05}$Mo$_{0.95})_{0.71}$B$_{2}$ displayed a critical field ($H_{\mrm{c}2}$) of 6.7 T, approaching the value of 9.4 T observed in MoB$_{2}$ at 110 GPa~\cite{Pei2023MoB2}. However, its maximum $\Tc$ was 
only
7.9 K, considerably lower than the high-pressure measurement of 32 K in MoB$_{2}$. In the Nb-substituted system, Nb-doping of 25\% yielded the highest ambient pressure $\Tc$ (onset) and $H_{\mrm{c}2}$ at 8.15 K and 6.7 T, respectively~\cite{Hire2022}. Our subsequent high-pressure study~\cite{Lim2023suppression} on this system showed that $\Tc$ \textit{decreases} from 8 K to 4 K between 0 to 50 GPa, followed by a steady yet subtle climb to 5 K at 171 GPa. 

In this paper, we report and compare results on the superconductivity in MoB$_{2}$-based systems using several approaches. 
The first approach further explores TM substitution in arc-melted compositions of the form (Mo$_{1-y}M_{y})_{x}$B$_{2}$ where $M$ is Zr, Hf, and Ta (Section~\ref{sec:arc-melt-results}). 
All of these alloys are stable in the P$6/mmm$ [191] phase with  (Mo$_{0.96}M_{0.04})_{0.85}$B$_{2}$ yielding the highest $\Tc$ at 8.60 K. 
The second approach is to decrease the cooling time during sample preparation. 
In this direction, we synthesized the TM-doped compositions mentioned above, as well as Nb$_{0.25}$Mo$_{0.75}$B$_{2}$, MoB$_{2}$, and MoB$_{2.5}$, using rapid cooling/quenching techniques (Section~\ref{sec:rapid-cool-results}). 
Details of our water-cooled splat-quenching procedure and apparatus can be found in the Methods Section~\ref{sec:methods}. Surprisingly, the rapidly cooled/quenched MoB$_{2}$ samples superconduct at ambient pressure with $\Tc$'s near 4.5 K. 
This is the first observation of superconductivity in MoB$_{2}$ at ambient pressure, possibly enabled by the creation of P6$/mmm$-like defects during the rapid cooling process. 


\section{Methods}\label{sec:methods}

\subsection{Sample Preparation and Characterization}
For experimental measurements, (Ta/Zr/Hf)$_{1-x}$Mo$_{x}$B$_{2}$ ($x = 0.04$, 0.10, 0.25, 0.4, 0.5) samples were formed via arc melting together the constituent elements. 
Mo foil (thickness 0.1 mm 99.97\% from AESAR) was used to wrap the other elements. Otherwise, boron (an insulator) sometimes breaks into small pieces when heated by the plasma arc.
A reasonable estimate for the temperature range for arc melting the constituent elements is between 2400$\celsius$ and 2700$\celsius$. 
Despite the high melting point of Mo (2622$\celsius$), the low vapor pressures of both B and Ta/Zr/Hf at this temperature led to negligible mass loss upon melting the constituents together, remelting twice.

Resistivity samples were cut from an arc-melted button using a low-speed diamond saw to dimensions of approximately $0.5\times0.5\times0.6$ mm$^{3}$. 
The sample was roughly rectangular with uniform thickness for the measurements. Small-scale errors arising from these assumptions were not taken into consideration. 
A current not exceeding 0.1 mA was used for all resistive measurements on the samples.
In a separate set of measurements, resistivity bars ($\rho$-bars) were made using a water-cooled caster to create a uniform thin bar with uniform dimensions of approximately $1\times1\times4$ mm$^{3}$. 
These $\rho$-bars were cooled faster ($\sim 10^{4}\,\celsius/\mrm{s}$) than the arc-melted button ($\sim 10\,\celsius/\mrm{s}$), which could take minutes to cool down from the melting point (e.g., 2000 $\celsius$). 
Resistivity measurements were done using the standard four-point probe method using a Keithley 220 programmable current source and a Keithley 2001 multimeter.
Specific heat at low temperatures was measured using a standard time constant methodology~\cite{Stewart1983}.

X-ray diffraction (XRD) measurements were conducted using a Panalytical XPert Powder system to identify the phases present in our crystalline sample. 
The material was initially fragmented into small pieces before being finely ground to ensure homogeneity. After measurement, the observed XRD pattern was cross-referenced with calculated patterns from the Materials Project database~\cite{MatProj2013} for accurate phase identification.

\subsection{Splat quenching}
%
\begin{figure}
    \includegraphics[trim={10.0cm 7.0cm 10.0cm 7.0cm}, clip, width=\linewidth]{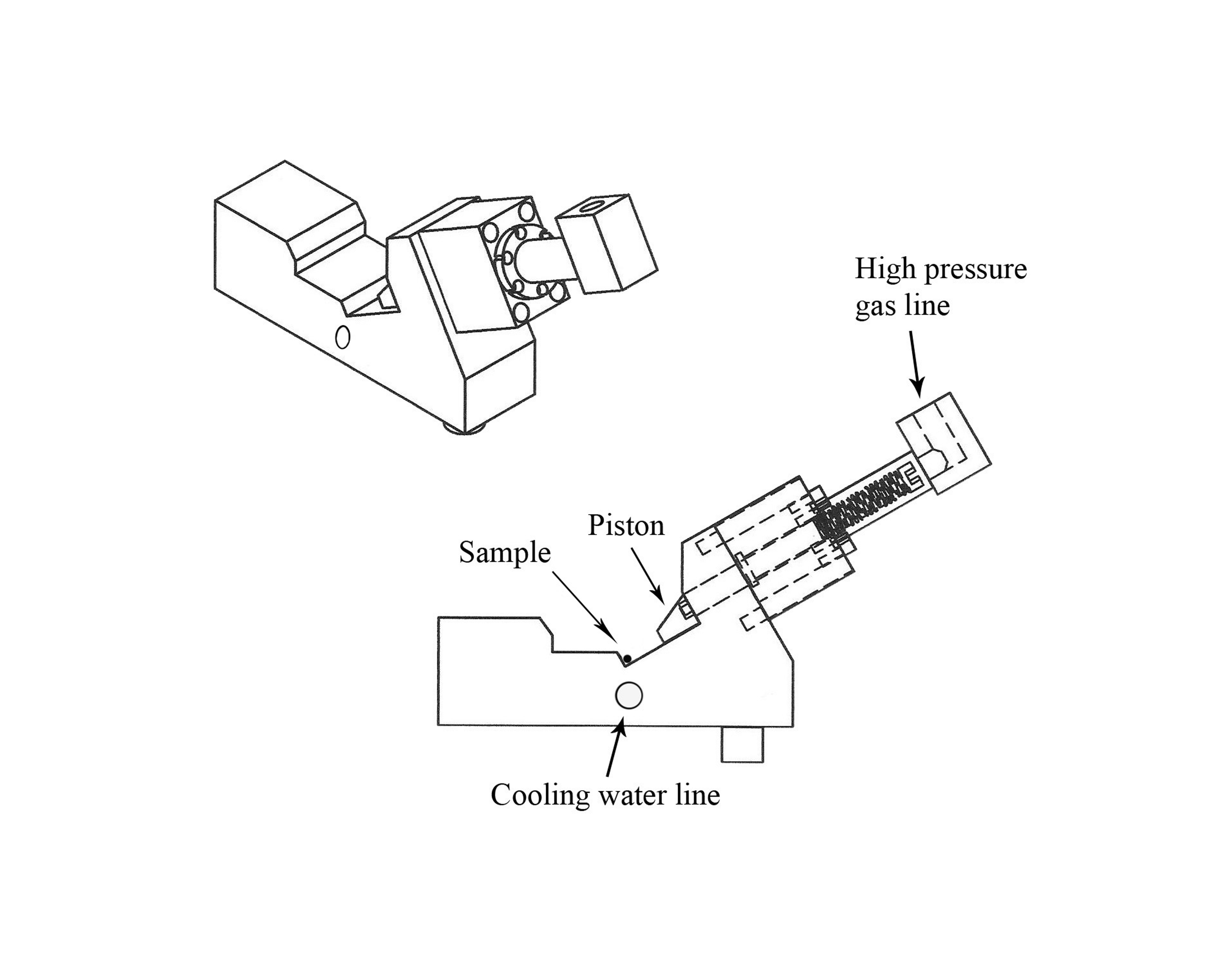}
    \caption{Schematic design of the splat-quench device. A high-pressure gas line is fed to a solenoid valve controlling a piston. The sample is seated at the base of the copper block, just above a channel fed by a water line for rapid cooling.   
    }
    \label{fig:Splat-quencher-diagram}
\end{figure}

\begin{figure}
    \centering
    \begin{adjustbox}{bgcolor=white}
    \includegraphics[trim={0.3cm 0.6cm 1.0cm 0.8cm}, clip, width=\linewidth]{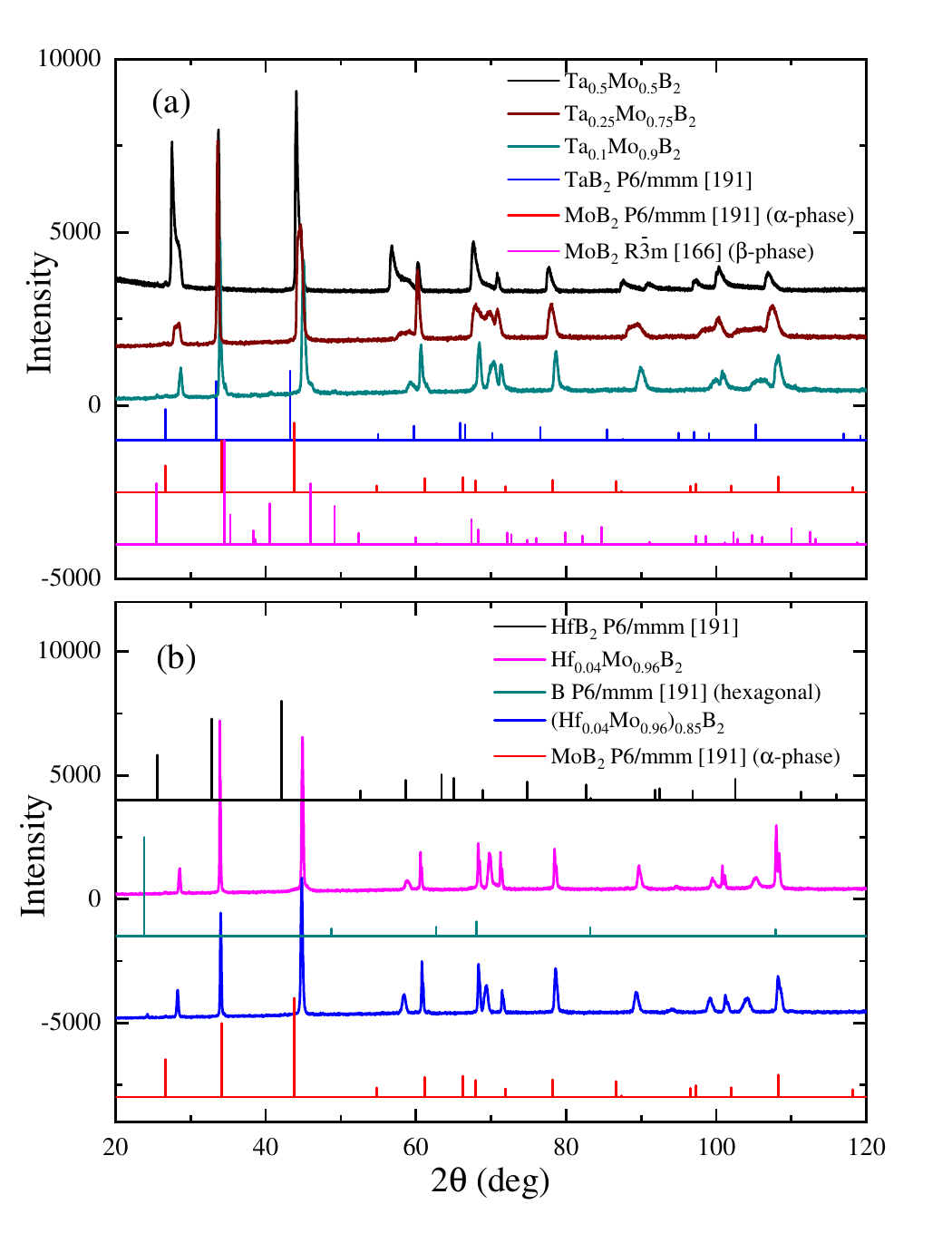}
    \end{adjustbox}
    \caption{X-ray diffraction measurements for (a) Ta- and (b) Hf-substituted MoB$_{2}$ prepared by arc melting synthesis procedure. We have included several relevant theoretical XRD results for easier comparison.
    }
    \label{fig:Ta-and-Hf-doped-MoB2_XRD}
\end{figure}
%
We designed and constructed a splat-quenching device (see Figure~\ref{fig:Splat-quencher-diagram}) to rapidly cool thin foil samples. This device consists of a copper block with a cooling water tube running through the block below the sample. It is securely affixed to the copper hearth of our arc-melter, ensuring stability during the quenching process. The design facilitates the close positioning of the arc-melter tip to the sample without interfering with other device components.  

The thin foil samples were produced by momentarily pressing a molten specimen using a piston arm powered by high-pressure argon gas (200 psi) through a solenoid valve. We utilized small samples (with a diameter of less than 1mm in their molten state) to ensure efficient melting and rapid cooling.

\begin{figure}
    \centering
    \begin{adjustbox}{bgcolor=white}
    \includegraphics[trim={0.6cm 1.2cm 0.9cm 0.7cm}, clip, width=0.96\linewidth]{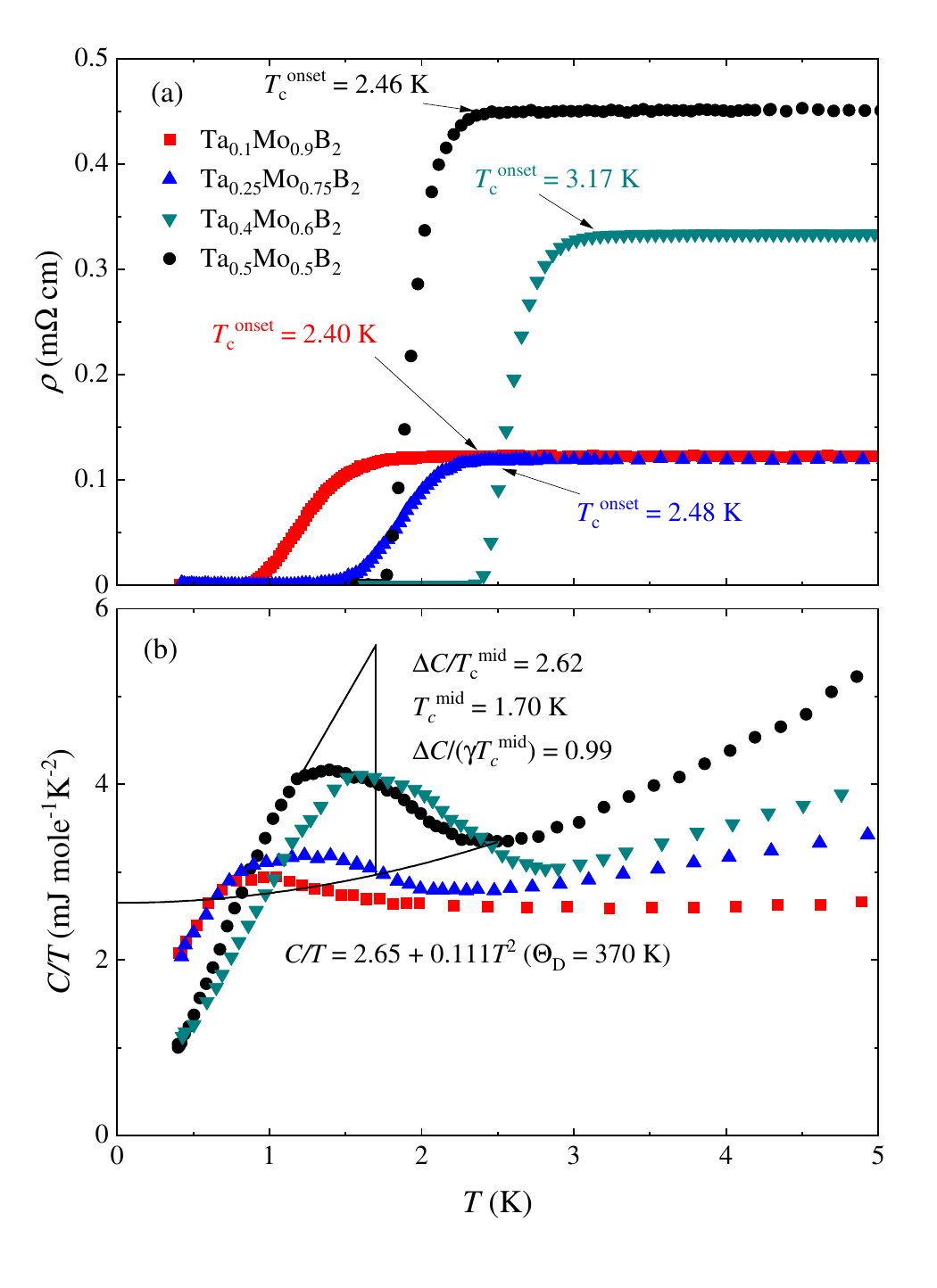}
    \end{adjustbox}
    \caption{Experimental results for the Ta$_{1-x}$Mo$_{x}$B$_{2}$ system for $x=0.1$, 0.25, 0.4, and 0.5. Panel (a) shows the resistivity measurements in m$\Omega$ cm, and (b) shows the specific heat per unit temperature in mJ mole$^{-1}$ K$^{-2}$, for each composition. In panel (b), a black line is included to show an example of the Debye model fit obtained for the $x=0.5$ composition. 
    }
    \label{fig:TaMoB2_rho_and_C_vs_T}
\end{figure}

\begin{figure}
    \centering
    \begin{adjustbox}{bgcolor=white}
    \includegraphics[trim={0.65cm 1.3cm 0.7cm 0.7cm}, clip, width=\linewidth]{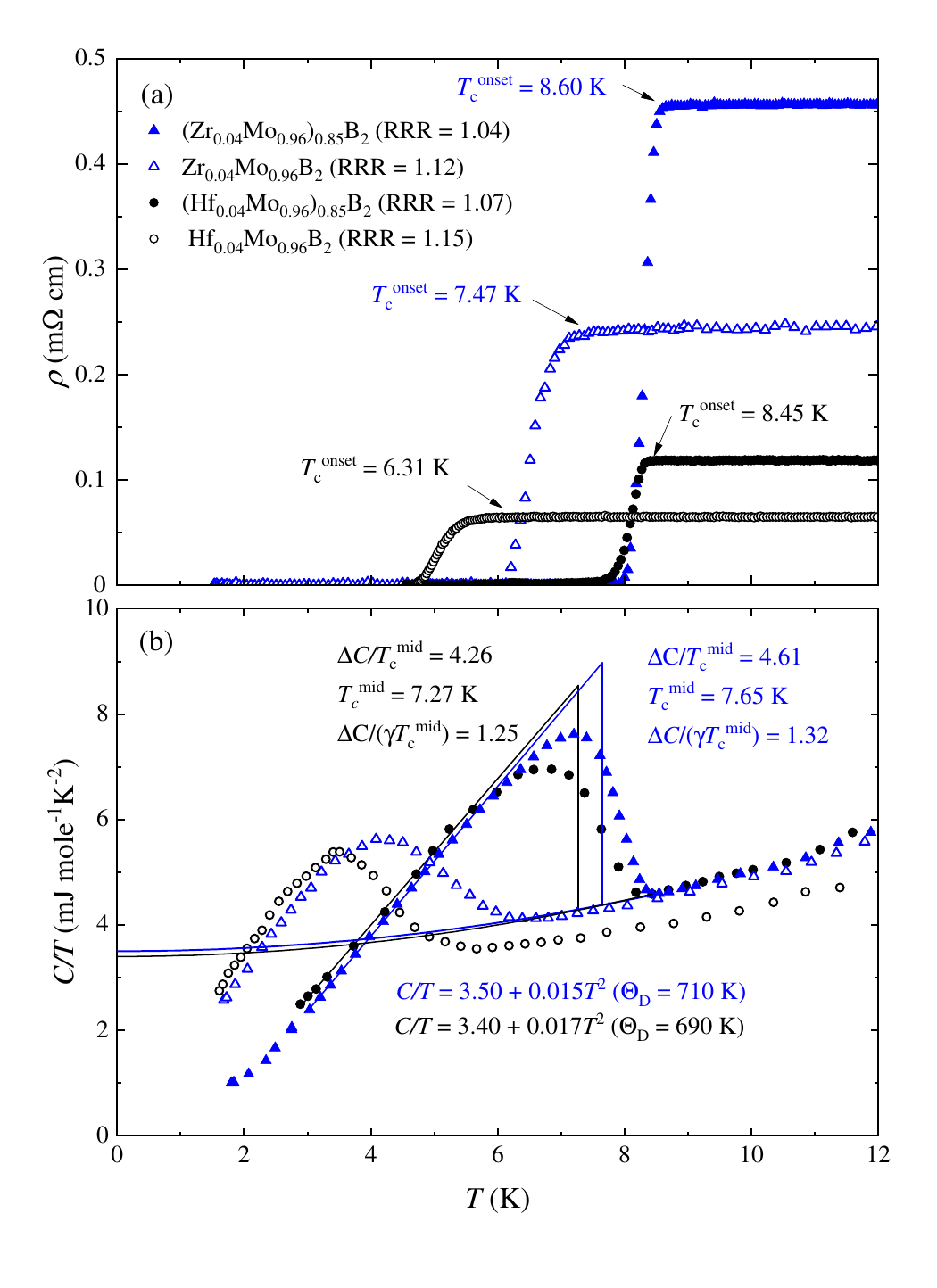}
    \end{adjustbox}
    \caption{Experimental results for the (Zr$_{0.04}$Mo$_{0.96})_{y}$B$_{2}$ and (Hf$_{0.04}$Mo$_{0.96})_{y}$B$_{2}$ systems for $y=0.85$ and 1.00.  Panel (a) shows the resistivity measurements in m$\Omega$ cm, and (b) shows the specific heat per unit temperature in mJ mole$^{-1}$ K$^{-2}$, for each composition. In panel (b), black and blue lines represent the Debye model fit obtained for the $y=0.85$ samples. 
    }
    \label{fig:Hf-Zr_4pcnt_doped_MoB2_rho_and_C_vs_T}
\end{figure}

\section{Results for arc-melted samples}\label{sec:arc-melt-results}

\subsection{X-ray diffraction}

Figure \ref{fig:Ta-and-Hf-doped-MoB2_XRD} shows the XRD results for the Ta- and Hf-doped MoB$_{2}$ compounds. Several relevant theoretical XRD patterns are shown for better comparison. These samples are best characterized as having the P$6/mmm$ [191] space group structure and are closer in alignment to the MoB$_{2}$ P$6/mmm$ [191] phase than that of TaB$_{2}$ or HfB$_{2}$. As expected, with higher Ta-substitution approaching 50\%, the XRD pattern shifts toward TaB$_{2}$. Additional XRD results for Zr-substituted are shown in the Supplemental Materials.

\subsection{Resistivity Measurements}
Resistivity and specific heat measurements were performed on all samples. 
The main results for Ta$_{x}$Mo$_{1-x}$B$_{2}$ with $x=0.1$, 0.25, 0.4, 0.5 are featured in Figure~\ref{fig:TaMoB2_rho_and_C_vs_T} while the remaining results for (Zr$_{0.04}$Mo$_{0.96}$)$_{y}$B$_{2}$ and (Hf$_{0.04}$Mo$_{0.96}$)$_{y}$B$_{2}$ for $y=0.85$, 1.0 are shown in Figure~\ref{fig:Hf-Zr_4pcnt_doped_MoB2_rho_and_C_vs_T}.
Resistivity measurements for each of the Ta-doped samples are shown in Fig.~\ref{fig:TaMoB2_rho_and_C_vs_T}(a) up to 5 K. 
The onset temperature of each superconducting transition $\Tcon$ was determined from the initial drop of the resistivity and found to be 2.40 K, 2.48 K, 3.17 K, and 2.46 K for nominal dopings $x=0.1$, 0.25, 0.4, and 0.5, respectively.   
The sharpest transition relative to $\Tcon$ is seen in Ta$_{0.5}$Mo$_{0.5}$B$_{2}$ in contrast with 10\% and 25\% Ta-doped samples. 
These trends are in accordance with the specific heat jumps, as shown in Fig.~\ref{fig:TaMoB2_rho_and_C_vs_T}(b).
The residual resistivity ratios (RRR) $[R(300\,\mrm{K}/R(\Tcon)]$ for $x=0.1$, 0.25, 0.4, and 0.5 are 1.29, 1.13, 1.09, and 1.09, respectively. 
Compared with pure MoB$_{2}$, which has a RRR of 2.74, these values point to increased scattering caused by alloying with Ta~\cite{Hire2022}.

The (Zr$_{0.04}$Mo$_{0.96}$)$_{y}$B$_{2}$ and (Hf$_{0.04}$Mo$_{0.96}$)$_{y}$B$_{2}$ samples yielded notably higher $\Tc$'s than the Ta-substituted series with $\Tcon$'s ranging from 6.31 K to 8.60 K, with the latter belonging to (Zr$_{0.04}$Mo$_{0.96}$)$_{0.85}$B$_{2}$ with a RRR of 1.04. 
These transition temperatures are comparable to the Zr-doped and Hf-doped results reported in Ref.~\cite{Muzzy2002}.
The complete list of results for the RRR and $\Tcon$ is summarized in Table~\ref{table_1}. 

\subsection{Specific Heat Measurements}


We characterize the low-temperature specific heat data using a Debye model given by
\begin{equation}
\frac{C}{T} = \gamma + \beta T^{2},
\end{equation}
where $\gamma$ and $\beta$ represent the linear (electronic) and cubic (phonon-related) specific heat coefficients, respectively.
The fitting coefficients were obtained using an entropy matching approach, ensuring the integral of $C/T$  versus $T$ from 0 to $\Tcon$ in the superconducting state corresponds precisely to its integral in the normal state.
From our estimated $\beta$ value, we calculated the Debye temperature $\Theta_{\mrm{D}}=(12\pi^{4}NR/5\beta)^{1/3}$, where $N$ denotes the number of atoms per formula unit and $R$ is the universal gas constant.
We determined the linear specific heat coefficient $\gamma=\lim_{T\rightarrow 0}C_{\mrm{normal}}/T$, which is proportional to the renormalized electronic density of states (DOS) at the Fermi energy, denoted as $N^{\ast}(0)$.

Applying Landau Fermi-liquid theory~\cite{Coleman_2015}, we approximated the experimental (renormalized) linear specific heat coefficient as $\gamma=\gamma^{(0)}(1+\lambda_{m})$, where $\gamma^{(0)}$ is the noninteracting case's linear coefficient, and $\lambda_{m} $ captures the electron mass enhancement factor near the Fermi level~\footnote{Formally, $\lambda_{m} := -\D\Sigma(\omega)/\D\omega|_{\omega=0}$ where $\Sigma(\omega)$ is the electron self-energy averaged over the Fermi surface $\Sigma(\omega):=\la \Sigma(\bk,\omega) \ra_{\bk\in\mrm{FS}}$.}
In principle, $\lambda_{m}$ includes effects from more than just electron-phonon interactions, but in our work, we assume the other effects to be small.
Hence, we take $\lambda_{m}\approx \lambda$ to estimate the renormalized and bare DOS, $N(0)$, where $\lambda$ is the electron-phonon coupling constant defined later.
Using this approximation, the relationship between the renormalized and bare electronic DOS, $\gamma$, and $\lambda$ is given as
\begin{equation}
    \gamma = \frac{\pi^{2}k_{\mrm{B}}^{2}}{3}N^{\ast}(0) = \frac{\pi^{2}k_{\mrm{B}}^{2}}{3}N(0)(1 + \lambda).
\end{equation}
We report these values in the last column of Table~\ref{table_1}.

We gauged the bulk superconductivity by the ratio $\Delta C/(\gamma\Tcmid)$, which equals a value of 1.43 in BCS superconductivity and, e.g., 1.65 in the unconventional, iron-based superconductor FeSe with $\Tc=8.1$ K~\cite{Lin2011}. 
Here, $\Tcmid$ is obtained from the peak of the entropy-matched Debye fit on the specific heat data (see Table~\ref{table_1}).
Assuming that $\Delta C/(\gamma\Tcmid)\approx 1.5$ indicates 100\% bulk superconductivity, the Ta-doped series shows 35\% to 71\% bulk superconductivity, similar to Nb-doped MoB$_{2}$~\cite{Hire2022}. 
In contrast, the (Zr$_{0.04}$Mo$_{0.96}$)$_{y}$B$_{2}$ and (Hf$_{0.04}$Mo$_{0.96}$)$_{y}$B$_{2}$ compositions exhibit 67\% to 88\% and 65\% to 83\% bulk superconductivity, respectively. 
We list the ratios and values for $\gamma$, $\beta$, and $\Theta_{\mrm{D}}$ for each composition in Table~\ref{table_1}.

\begin{table*}
\begin{tabular}{lcrrrrcccccccc}
\hline 
\multirow{2}{*}{Material} & \mcolc{\multirow{2}{*}{Synthesis}} & \mcolc{$\Tc^{\text{onset}}$} & \mcolc{$\Tc^{\text{mid}}$} & \mcolc{\multirow{2}{*}{$\frac{\Delta C}{\gamma \Tcmid}$}} & \mcolc{\multirow{2}{*}{RRR}} &  \multicolumn{2}{c}{\footnotesize Measured [\AA]} &  \mcolc{$\gamma$}  & \mcolc{$\beta$} & \mcolc{$\Theta_{\mrm{D}}$} & \mcolc{$\lambda$} & \multicolumn{2}{c}{{\footnotesize[states/eV/f.u.]}} \\
                     &    & \mcolc{\footnotesize [K]} & \mcolc{\footnotesize [K]}   &   &   & \mcolc{$a$} & \mcolc{$c$} & \mcolc{\footnotesize [mJ/mol K$^2$]}  & \mcolc{\footnotesize [mJ/mol K$^4$]} & \mcolc{\footnotesize [K]} &  & \mcolc{$N^{\ast}(0)$} & \mcolc{$N(0)$}  \\
\hline \hline
Ta$_{0.1}$Mo$_{0.9}$B$_{2}$   				&  a.m.        & 2.40  & 1.28  & 0.53  & 1.29 & 3.051 & 3.341 &  2.50 & 0.010 & 850  &  0.437 & 1.06 & 0.74 \\
Ta$_{0.25}$Mo$_{0.75}$B$_{2}$ 				&  a.m.        & 2.48  & 1.42  & 0.64  & 1.13 & 3.081 & 3.308 &  2.47 & 0.052 & 480  &  0.485 & 1.05 & 0.71 \\
Ta$_{0.4}$Mo$_{0.6}$B$_{2}$   				&  a.m.        & 3.17  & 1.93  & 1.06  & 1.09 & 3.068 & 3.247 &  2.67 & 0.046 & 500  &  0.504 & 1.13 & 0.75 \\
Ta$_{0.5}$Mo$_{0.5}$B$_{2}$	  				&  a.m.        & 2.46  & 1.70  & 0.99  & 1.09 & 3.068 & 3.249 &  2.65 & 0.111 & 370  &  0.509 & 1.12 & 0.74 \\
Zr$_{0.04}$Mo$_{0.96}$B$_{2}$				&  a.m.        & 7.47  & 4.72  & 1.01  & 1.12 & 3.052 & 3.349 &  3.53 & 0.013 & 760  &  0.555 & 1.50 & 0.96 \\
(Zr$_{0.04}$Mo$_{0.96}$)$_{0.85}$B$_{2}$	&  a.m.        & 8.60  & 7.65  & 1.32  & 1.04 & 3.064 & 3.371 &  3.50 & 0.015 & 710  &  0.585 & 1.48 & 0.94 \\
(Zr$_{0.04}$Mo$_{0.96}$)$_{0.85}$B$_{2}$	&  $\rho$-bar  & 9.60  & 7.09  & 1.09  & 1.07 &\cdash &\cdash &  3.12 & 0.017 & 690  &  0.607 & 1.32 & 0.82 \\
(Zr$_{0.04}$Mo$_{0.96}$)$_{0.85}$B$_{2}$	&  w.c.s.q.    & 10.14 &\cdash &\cdash & 1.08 &\cdash &\cdash &\cdash &\cdash &\cdash&  \cdash&\cdash&\cdash\\
Hf$_{0.04}$Mo$_{0.96}$B$_{2}$				&  a.m.        & 6.31  & 4.07  & 0.97  & 1.15 & 3.052 & 3.344 &  3.35 & 0.006 & 970  &  0.507 & 1.42 & 0.94 \\
(Hf$_{0.04}$Mo$_{0.96}$)$_{0.85}$B$_{2}$	&  a.m.        & 8.45  & 7.27  & 1.25  & 1.07 & 3.045 & 3.345 &  3.40 & 0.017 & 690  &  0.587 & 1.44 & 0.91 \\
MoB$_{2}$ ($R\bar{3}m)$       				&  a.m.        &$<1.7$ &$<1.7$ &\cdash & 2.74 &\cdash &\cdash &\cdash &\cdash &\cdash&\cdash  &\cdash&\cdash\\
MoB$_{2}$                     				&  $\rho$-bar  & 4.45  & 4.08  & 0.76  & 1.23 &\cdash &\cdash &  5.11 & 0.135 & 350  & 0.592  & 2.17 & 1.36 \\
MoB$_{2}$                     				&  w.c.s.q.    & 4.55  &\cdash &\cdash & 1.28 &\cdash &\cdash &\cdash &\cdash &\cdash&\cdash  &\cdash&\cdash\\
MoB$_{2.5}$                     			&  a.m.       & 3.06  &\cdash &\cdash & 1.38 &\cdash &\cdash &\cdash &\cdash &\cdash&\cdash  &\cdash&\cdash\\
MoB$_{2.5}$                     			&  $\rho$-bar  & 5.82  & 2.88  & 0.76  & 1.10 &\cdash &\cdash & 3.36  & 0.130 & 810 & 0.517 & 1.43 & 0.94 \\
Nb$_{0.25}$Mo$_{0.75}$B$_{2}$           	&  a.m.        & 8.05  & 6.84  & 1.00  & 1.07 & 3.055\Hire & 3.264\Hire & 3.79  & 0.014 & 740  & 0.569  & 1.61 & 1.02 \\
Nb$_{0.25}$Mo$_{0.75}$B$_{2}$      			&  $\rho$-bar  & 10.67 & 8.44  & 0.89  & 1.10 &\cdash &\cdash & 3.94  & 0.016 & 710  & 0.620  & 1.67 & 1.03 \\
Nb$_{0.25}$Mo$_{0.75}$B$_{2}$      			&  w.c.s.q.    & 10.45 &\cdash &\cdash & 1.07 &\cdash &\cdash &\cdash &\cdash &\cdash&\cdash  &\cdash&\cdash\\

\hline\hline
\end{tabular}
\caption{Summary of experimental results for ([Ta/Zr/Hf/Nb]$_{y}$Mo$_{1-y}$)$_{x}$B$_{2}$ in the P6/$mmm$ phase and for MoB$_{2}$ in the R$\bar{3}m$ phase. In the arc-melted MoB$_{2}$, no superconductivity was observed down to 1.7 K, consistent with the literature~\cite{Cooper1970}. The correct space groups for the $\rho$-bar and water-cooled splat-quenched MoB$_{2}$ samples are likely to be R$\bar{3}m$ but remain unknown. The DOS is stated per eV per formula unit.
\\
a.m.: ``arc-melted''
\\
$\rho$-bar: created using $\rho$-bar cooling technique
\\
w.c.s.q.: ``water-cooled splat-quenched''}
\label{table_1}
\end{table*}
\subsection{Estimate of the electron-phonon coupling}
Equipped with $\Tc$ from the resistivity results and the Debye temperature from the specific heat fits, we estimate the electron-phonon coupling constant $\lambda$ using the inverted McMillan formula~\cite{McMillan1968}
\begin{equation}\label{eqn:lambda_McMill}
    \lambda = \frac{
        1.04 + \mu^{\ast} \ln[\Theta_{\mrm{D}}/(1.45\Tc)] 
    }{
        (1 - 0.62\mu^{\ast}) \ln[\Theta_{\mrm{D}}/(1.45\Tc)] - 1.04    
    },
\end{equation}
where $\mu^{\ast}$ is the Coulomb pseudopotential parameter. For better comparison, we follow Yang et al.~\cite{Yang2022} and Quan et al.~\cite{Quan2021} and take $\mu^{\ast}=0.13$. For our samples involving Ta-, Zr-, Hf, and Nb-substituted MoB$_{2}$, we report the estimates for $\lambda$ in the last column of Table~\ref{table_1}. The value of $\mu^{\ast}=0.13$ sits in the middle of the range 0.1 to 0.15, a standard reference range for diborides~\cite{Quan2021}. Using the limits of this range, the estimates for $\lambda$ will decrease by roughly 0.06 for $\mu^{\ast}=0.1$ and increase by 0.04 for $\mu^{\ast}=0.15$ as compared with our choice of $\mu^{\ast}=0.13$. 

Regardless of the specific choice of $\mu^{\ast}$, the procedure above indicates that, on the whole, nearly all the compounds studied in this work have $\lambda$ between 0.4 and 0.65, consistent with similar estimates on many TM-diborides~\cite{Shein2006,Heid2003,Yang2022}. This range of values is considered to be relatively weak by e-ph coupling standards.







\begin{figure*}
    \includegraphics[trim={0.2cm 0.2cm 0.1cm 0.2cm}, clip, width=0.95\linewidth]{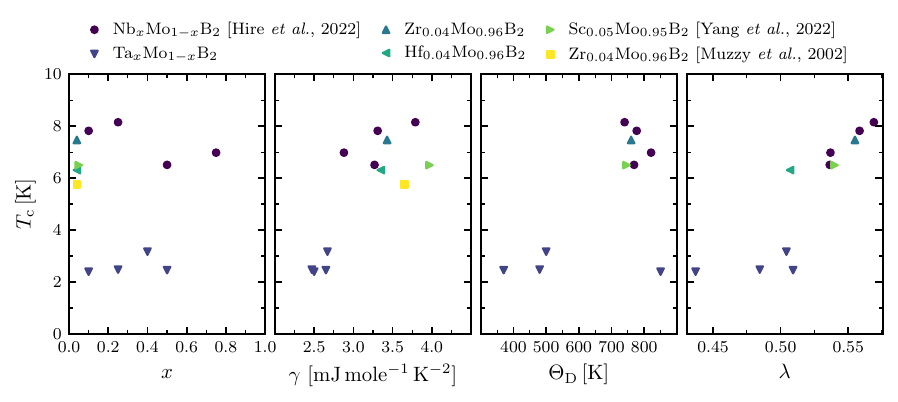}
    \caption{Correlations of $\Tc$ with various quantities for various TM-substituted MoB$_{2}$ samples. Literature values are included from Refs.~\cite{Hire2022} (Nb),~\cite{Muzzy2002} (Zr), and~\cite{Yang2022} (Sc). In Ref.~\cite{Yang2022}, the authors report $\Tcmid = 4.62$ K instead of $\Tcon$. To be consistent, we've plotted an estimate of $\Tcon\approx 6.5$ K from their resistivity data.}
    \label{fig:diboride_compare}
\end{figure*}

\subsection{Comparison with other alloyed MoB$_{2}$ compounds}
Incorporating our recent measurements on the Ta-, Zr-, and Hf-substituted compounds, alongside our earlier results on the Nb$_{x}$Mo$_{1-x}$B$_{2}$ system and the findings from Refs.~\cite{Muzzy2002,Yang2022} regarding Zr- and Sc-doped MoB$_{2}$ alloys, we have compiled a comprehensive summary of compositionally similar results to date. Utilizing the nominal composition and specific heat measurements, we sought potential correlations between the superconducting transition temperature $\Tc$ and the key properties of each system. These include the non-Mo TM-element doping $x$, linear specific heat contribution $\gamma$, Debye temperature $\Theta_{\mrm{D}}$, and the electron-phonon (e-ph) coupling strength $\lambda$, as illustrated in Fig.~\ref{fig:diboride_compare}.

Scrutiny of Fig.~\ref{fig:diboride_compare} reveals a weak correlation between $\Tc$ and the evaluated parameters, except for a modest association with $\lambda$. However, this modest correlation may arise from our empirical approach to estimating $\lambda$ using Eqn.~\ref{eqn:lambda_McMill} instead of a full ab initio evaluation. In our prior study on Nb$_{x}$Mo$_{1-x}$B$_{2}$, ab initio estimates of $\Tc$ via density functional theory and the Allen-Dynes formula~\cite{Allen-Dynes1975} consistently exceeded experimental results by a factor of 2 or more~\cite{Lim2023suppression}. We proposed potential sources of this discrepancy, such as sample inhomogeneity, perhaps best exemplified by the formation of vacancies and stacking faults in transition metal diborides~\cite{Muzzy2002}. Regardless of the underlying cause, our data strongly suggests that these compounds exhibit weak coupling superconductivity under ambient pressure conditions.

Noticeably absent from Fig.~\ref{fig:diboride_compare} are non-stoichiometric results where the ratio of boron to the TM atoms is greater than 2:1. We have included a separate comparison of $\Tc$'s among these materials in Table~\ref{table_2}. There, we have also included similar compositions from a few sources in the literature. 

\begin{figure}
    \centering
    \begin{adjustbox}{bgcolor=white}
    \includegraphics[trim={0.3cm 0.4cm 1.3cm 1.3cm}, clip, width=\linewidth]{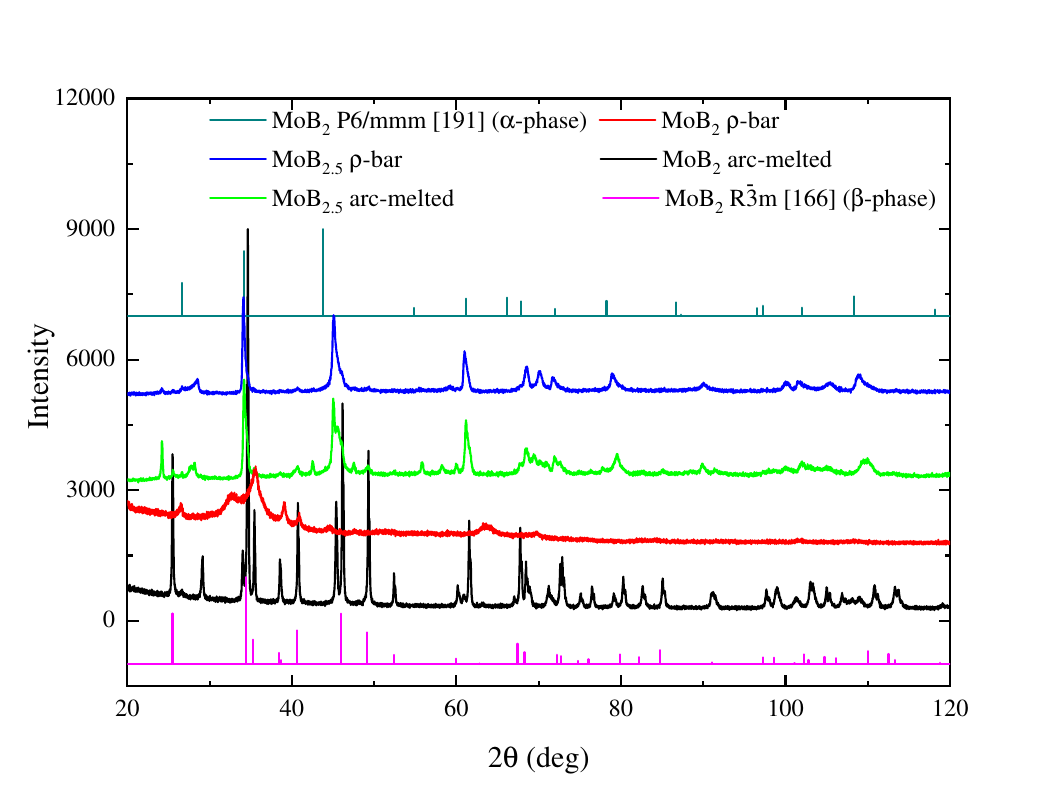}
    \end{adjustbox}
    \caption{X-ray diffraction measurements for MoB$_{2}$ and MoB$_{2.5}$ prepared by arc melting and the $\rho$-bar (rapid-cooling) synthesis procedure. The top and bottom diffraction patterns are the theoretical XRD results for MoB$_{2}$ in the P$6/mmm$ [191] and R$\bar{3}m$ [166] phases, respectively. 
    }
    \label{fig:MoB2_XRD}
\end{figure}

\begin{figure}
    \centering
    \begin{adjustbox}{bgcolor=white}
    \includegraphics[trim={0.2cm 0.4cm 1.1cm 0.7cm}, clip, width=\linewidth]{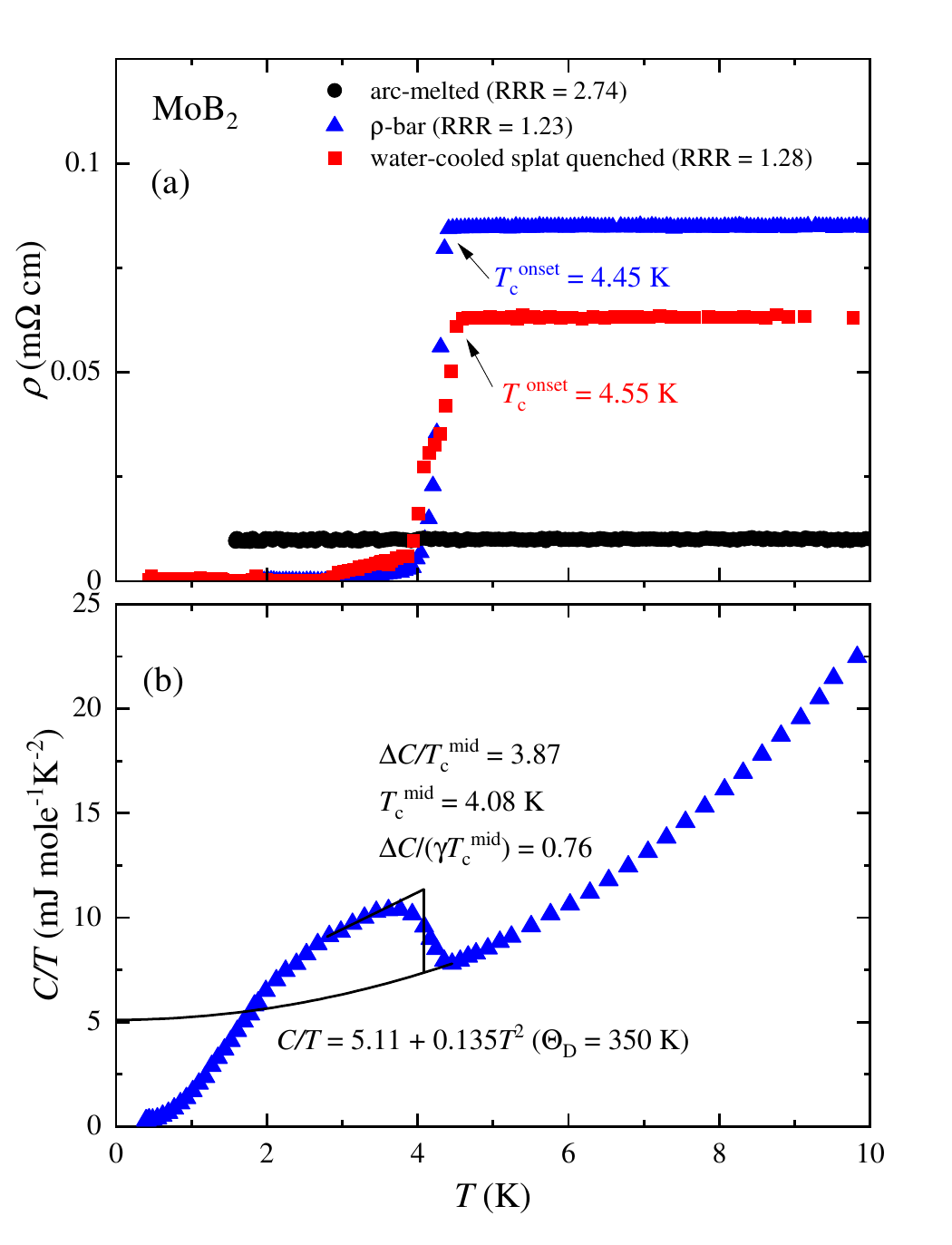}
    \end{adjustbox}
    \caption{Experimental results for MoB$_{2}$ showing the (a) resistivity $\rho$ in [m$\Omega$ cm] for the arc-melted, $\rho$-bar, and water-cooled splat-quenched samples, as well as (b) the specific heat per unit temperature $C/T$ in [mJ mole$^{-1}$K$^{-2}$] for the $\rho$-bar sample only.}
    \label{fig:MoB2_2x1}
\end{figure}
\begin{figure}
    \centering
    \begin{adjustbox}{bgcolor=white}
    \includegraphics[trim={0.4cm 0.6cm 1.3cm 0.7cm}, clip, width=\linewidth]{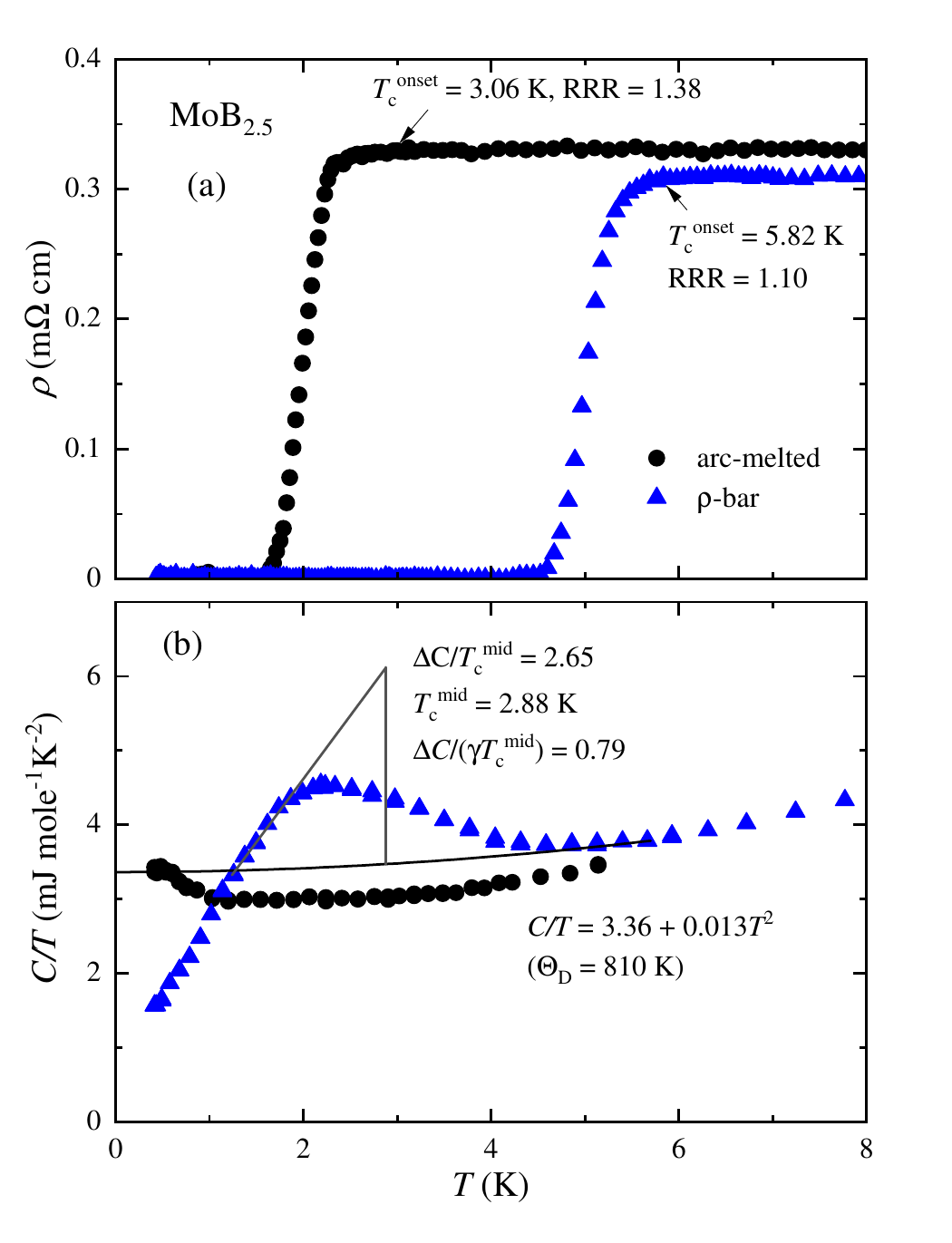}
    \end{adjustbox}
    \caption{
    Experimental results for MoB$_{2.5}$ showing the (a) resistivity $\rho$ in [m$\Omega$ cm], as well as (b) the specific heat per unit temperature $C/T$ in [mJ mole$^{-1}$K$^{-2}$] the arc-melted and $\rho$-bar samples.
    }
    \label{fig:MoB2p5_2x1}
\end{figure}
\section{Analysis of rapidly cooled samples}\label{sec:rapid-cool-results}
The early work by Cooper \textit{et al.}~\cite{Cooper1970} used splat-quenching to make MoB$_{2}$, as typical arc melting seemed to produce a slightly B-deficient sample. However, their quenched MoB$_{2}$ sample did not exhibit superconductivity down to the lowest temperature measured, 1.8 K. Only for a splat-quenched sample of MoB$_{2.5}$ did they observe a transition ($\Tc=8.1$ K). We have reexamined the potential of rapid cooling during synthesis to generate favorable conditions for superconductivity in MoB$_{2}$ at \textit{ambient} pressure. Surprisingly, both rapidly cooled MoB$_{2}$ samples exhibited superconductivity. As measured from the initial drop in the resistivity, the $\rho$-bar and water-cooled splat-quenched samples had $\Tc$'s of 4.45 K and 4.55 K, respectively [see Fig.~\ref{fig:MoB2_2x1}(a)]. We measured the specific heat of the $\rho$-bar sample, and the result is plotted in Fig.~\ref{fig:MoB2_2x1}(b). The specific heat jump is broad, and the Debye entropy-matching procedure yields $\gamma = 5.11\,\mrm{mJ\,mol^{-1}\,K^{-2}}$, and $\Theta_{\mrm{D}}=350$ K. With $\Delta C/(\gamma\Tcmid)=0.76$, we can assume that around 50\% of the sample is superconducting. This may be the first observation of superconductivity in stoichiometric ambient pressure MoB$_{2}$ if excess boron can be ruled out. To better distinguish our MoB$_{2}$ from an excess boron phase, we also synthesized arc-melted and $\rho$-bar samples of MoB$_{2.5}$. 


Figure~\ref{fig:MoB2_XRD} shows the XRD results for MoB$_{2}$ and MoB$_{2.5}$. The top and bottom XRD patterns are the theoretical results for MoB$_{2}$ in the P$6/mmm$ (``$\alpha$-phase'') and $R\bar{3}m$  (``$\beta$-phase''), respectively. The arc-melted MoB$_{2}$ sample is better aligned with the $R\bar{3}m$ [166] structure, as expected~\cite{Pei2023MoB2,Quan2021}.
The XRD results for the $\rho$-bar MoB$_{2}$ sample are far less conclusive. Unfortunately, the ill-defined peaks of the $\rho$-bar XRD pattern make it difficult to assign a structural phase. A few discernible peaks align somewhat with the R$\bar{3}m$ phase, but the lowest angle peak better matches that of the P$6/mmm$ phase. This may indicate a mixture of R$\bar{3}m$ and P$6/mmm$-like defects analogous to the pressure-induced stacking faults in WB$_{2}$~\cite{Lim2022}. The lack of a clear match in the XRD results implies that the precise stoichiometry is uncertain. Our nominal composition is based solely on the ratios of the constituents used in the arc-melting process. In comparison, the XRD results for MoB$_{2.5}$ have well-defined peaks and reveal that the arc-melted and $\rho$-bar samples are in the P$6/mmm$ [191] phase. These results are quite distinct from those of $\rho$-bar MoB$_{2}$, helping to rule out excess boron in the latter.

Nevertheless, using Eqn.~(\ref{eqn:lambda_McMill}), we estimated the electron-phonon coupling constant in our $\rho$-bar sample MoB$_{2}$ to be $\lambda \approx 0.59$. 
This empirical value should be interpreted cautiously, especially when compared to recent ab initio calculations for the $\alpha$-phase under 90 GPa pressure, which suggest a $\lambda$ range of approximately 1.6 to 1.71~\cite{Pei2023MoB2,Quan2021,Liu2022}. 
These sources do not report a value for $\lambda$ in the lower pressure $\beta$-phase. However,  Ref.~\cite{Pei2023MoB2} reports a theoretical $\Tc$ of 5 K, implying $\lambda$ is likely much smaller than that of the $\alpha$-phase under pressure. They ascribe the low $\Tc$ to a reduced electronic DOS at the Fermi level.
The high-pressure phase is predicted to feature a larger electronic DOS near the Fermi level, thanks mostly to the presence of van Hove singularities attributed to Mo $d_{z^{2}}$ bands~\cite{Liu2022}. 
However, the high-pressure phase is also predicted to have a boosted contribution from boron $p$ bands, which may have an outsized effect on increasing $\Tc$~\cite{Quan2021}. 
Using the experimental results for the $\rho$-bar MoB$_{2}$, we estimate the renormalized (bare) DOS near the Fermi level to be $N^{\ast}(0)\approx 2.17$ states/eV/f.u. ($N(0)\approx 1.36$ states/eV/f.u.). 

In Figure~\ref{fig:MoB2p5_2x1}(a), we show our results for the resistivity of each MoB$_{2.5}$ sample. 
These arc-melted and $\rho$-bar samples have a broader transition than the $\rho$-bar MoB$_{2}$ with $\Tcon=3.06$ and 5.82 K, respectively. Moreover, the MoB$_{2.5}$ samples have higher resistivity above $\Tcon$ and lower RRR values, suggesting they contain more defects. Without further analysis, it is difficult to characterize the nature of these defects, though the possible phases and role and whereabouts of the excess boron have been argued in previous studies~\cite{Burdett1986,Klesnar1996}.

The specific heat measurements of the arc-melted and $\rho$-bar MoB$_{2.5}$ samples are shown in Figure~\ref{fig:MoB2p5_2x1}(b). The specific heat peak of the arc-melted samples (black circles) was not fully resolvable down to 0.41 K, and the addenda contribution approaches 50\% at the highest temperature of 5.2 K. The $\rho$-bar samples (blue triangles) show a fairly broad peak, and the entropy matching procedure places $\Tcmid$ at 2.88 K, considerably lower than the main drop in resistivity. The Debye coefficients are $\gamma=3.36\, \mrm{mJ\,mol^{-1}\,K^{-2}}$ and $\beta=0.013\,\mrm{mJ\,mol^{-1}\,K^{-4}}$, the latter leading to a Debye temperature of $\Theta_{\mrm{D}}=810\,\mrm{K}$. These results lead to a slightly weaker $\lambda\sim0.52$ and DOS at the Fermi level as compared with the MoB$_{2}$ $\rho$-bar samples (see Table~\ref{table_1} ).


\begin{figure}
    \centering
    \begin{adjustbox}{bgcolor=white}
    \includegraphics[trim={0.4cm 0.35cm 1.2cm 0.9cm}, clip, width=\linewidth]{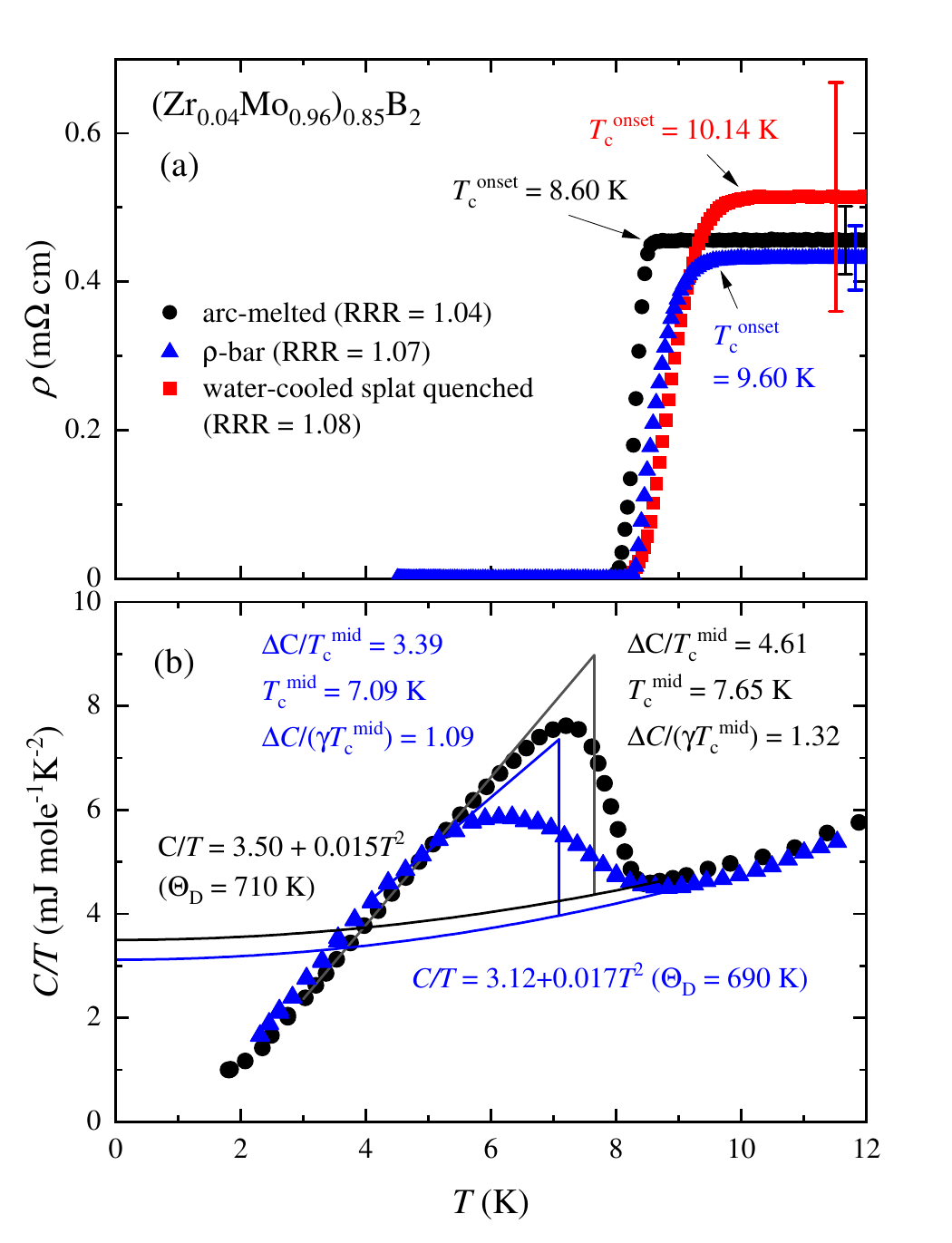}
    \end{adjustbox}
    \caption{Experimental results for (Zr$_{0.04}$Mo$_{0.96})_{0.85}$B$_{2}$ showing the (a) resistivity $\rho$ in [m$\Omega$ cm] for the arc-melted, $\rho$-bar, and water-cooled splat-quenched samples, as well as (b) the specific heat per unit temperature $C/T$ in [mJ mole$^{-1}$K$^{-2}$] for the arc-melted and $\rho$-bar samples.
    }
    \label{fig:(Zr0.04Mo0.96)_0.85B2_arc-rhobar-splat}
\end{figure}

\begin{figure}
    \centering
    \begin{adjustbox}{bgcolor=white}
    \includegraphics[trim={0.4cm 0.6cm 1.2cm 0.8cm}, clip, width=\linewidth]{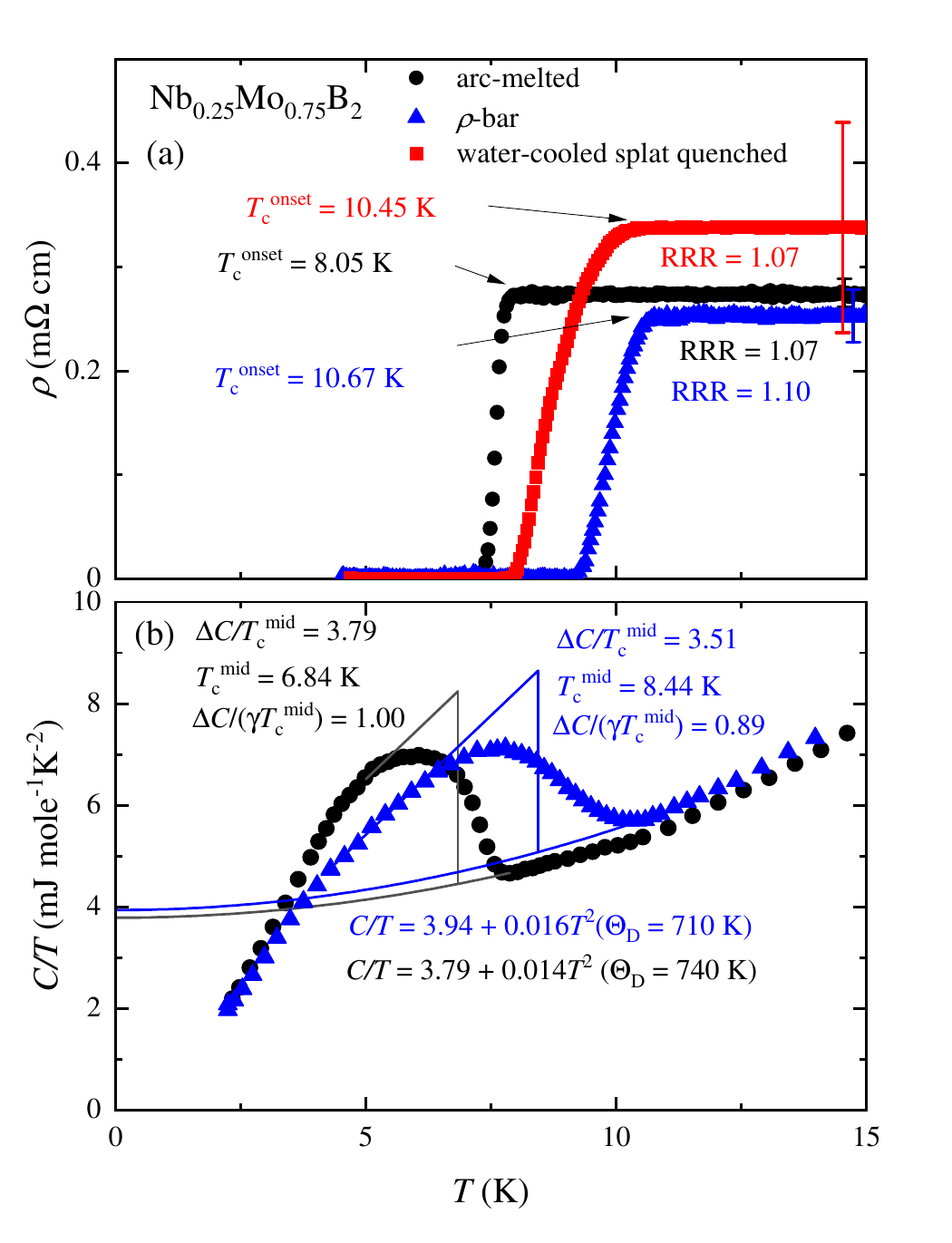}
    \end{adjustbox}
    \caption{Experimental results for Nb$_{0.25}$Mo$_{0.75}$B$_{2}$ showing the (a) resistivity $\rho$ in [m$\Omega$ cm] for the arc-melted, $\rho$-bar, and water-cooled splat-quenched samples, as well as (b) the specific heat per unit temperature $C/T$ in [mJ mole$^{-1}$K$^{-2}$] for the arc-melted and $\rho$-bar samples.
    }
    \label{fig:Nb0.25Mo0.75B2_2x1}
\end{figure}


Turning our focus to the (Zr$_{0.04}$Mo$_{0.96})_{0.85}$B$_{2}$ composition, which exhibited the highest transition temperature ($\Tc=8.60$ K) of the arc-melted samples, we undertook further synthesis using the $\rho$-bar and water-cooled splat-quenching methods. The comparative results are shown in Figure~\ref{fig:(Zr0.04Mo0.96)_0.85B2_arc-rhobar-splat}. The water-cooled splat-quenched sample yielded a slightly higher $\Tcon$ of 10.14 K compared with the $\rho$-bar sample $\Tcon = 9.60$ K [Fig.~\ref{fig:(Zr0.04Mo0.96)_0.85B2_arc-rhobar-splat}(a)]. However, both have broader resistivity drops as compared with the arc-melted sample. This trend is further exemplified by comparing the specific heat jumps of the arc-melted and $\rho$-bar samples in Fig.~\ref{fig:(Zr0.04Mo0.96)_0.85B2_arc-rhobar-splat}(b). The arc-melted sample displays more bulk superconductivity, with a $\Delta C/ (\gamma\Tcmid)\sim 1.32$, compared with 1.09 in the $\rho$-bar system. We determine $\Theta_{\mrm{D}}=710$ and 690 K for the arc-melted and $\rho$-bar from the Debye fitting procedure. Our estimate for the e-ph coupling is $\lambda\approx 0.61$, the largest estimate obtained thus far, albeit relatively weak.    

We synthesized $\rho$-bar and water-cooled splat-quenched samples of Nb$_{0.25}$Mo$_{0.75}$B$_{2}$ and plotted their resistivity curves in Fig.~\ref{fig:Nb0.25Mo0.75B2_2x1}(a). These curves reveal a considerable spread among the samples. The transition for the arc-melted sample starts at $\Tcon=8.05$ K, followed by the water-cooled splat-quenched sample with $\Tcon=10.45$ K, and topped by the $\rho$-bar sample with $\Tc=10.67$. These samples exhibit similar RRR values, ranging from 1.07 to 1.10.
Compared to the arc-melted sample, the $\rho$-bar and water-cooled splat-quenched samples show broader transitions. This observation is most evident when comparing their specific heat measurements, as shown in Fig.~\ref{fig:Nb0.25Mo0.75B2_2x1}(b). The $\rho$-bar sample exhibits a lower $\Delta C/ (\gamma\Tcmid)$ ratio of 0.89, as opposed to 1.00 for the arc-melted sample.
The linear specific heat coefficients were $\gamma = 3.79$ and 3.94 $\mrm{mJ\,mol^{-1}\,K^{-2}}$ for the arc-melted and $\rho$-bar samples, respectively. Additionally, the Debye temperatures ($\Theta_{\mrm{D}}$) were found to be 740 K and 710 K for the arc-melted and $\rho$-bar samples, respectively.

\begin{table}
\begin{tabular}{lcrrc}
\hline 

 \multirow{2}{*}{Material} & \multicolumn{1}{c}{\multirow{2}{*}{Synthesis}} & \multicolumn{1}{c}{$\Tc^{\text{onset}}$} & \multicolumn{1}{c}{$\Tc^{\text{mid}}$} & \multicolumn{1}{c}{\multirow{2}{*}{Reference}} \\
                           &                            & \multicolumn{1}{c}{{\footnotesize [K]} } & \multicolumn{1}{c}{{\footnotesize [K]} } &                     \\
                          
\hline \hline
(Zr$_{0.04}$Mo$_{0.96}$)$_{0.85}$B$_{2}$ &  a.m.        & 8.60       & 7.65  &  This work                \\
(Zr$_{0.04}$Mo$_{0.96}$)$_{0.85}$B$_{2}$ &  $\rho$-bar  & 9.60       & 7.09  &  This work                \\
(Zr$_{0.04}$Mo$_{0.96}$)$_{0.85}$B$_{2}$ &  w.c.s.q.    & 10.14      & \cdash    &  This work                \\
(Zr$_{0.04}$Mo$_{0.96}$)$_{0.85}$B$_{2}$ &  a.m.        & 8.137      & \cdash    & \cite{Muzzy2002}  \\
(Hf$_{0.04}$Mo$_{0.96}$)$_{0.85}$B$_{2}$ &  a.m.        & 8.45       & 7.27  &  This work                \\
(Hf$_{0.04}$Mo$_{0.96}$)$_{0.80}$B$_{2}$ &  a.m.        & $\sim8.0$  & \cdash     & \cite{Muzzy2002}  \\
(Ta$_{0.04}$Mo$_{0.96}$)$_{0.80}$B$_{2}$ &  a.m.        & $<4.0$     & \cdash     & \cite{Muzzy2002}  \\
(W$_{0.04}$Mo$_{0.96}$)$_{0.80}$B$_{2}$  &  a.m.        & $<4.0$     & \cdash     & \cite{Muzzy2002}  \\
(Nb$_{0.95}$Mo$_{0.05}$)$_{0.80}$B$_{2}$ &  a.m.        & $\sim3.5$  & \cdash     & \cite{Muzzy2002}  \\
(Ti$_{0.04}$Mo$_{0.96}$)$_{0.80}$B$_{2}$ &  a.m.        & 5.0        & \cdash     & \cite{Muzzy2002}  \\
(Ti$_{0.04}$Mo$_{0.96}$)$_{0.80}$B$_{2}$ &  a.m.        & 7.4        & 7.0$^{\dagger}$   & \cite{Yang2023}   \\
(Sc$_{0.05}$Mo$_{0.95}$)$_{0.83}$B$_{2}$ &  a.m.        & 6.01       & \cdash     & \cite{Yang2022}   \\

\hline \hline
\end{tabular}
\caption{Survey of excess-boron compositions of the form ($M_{y}$Mo$_{1-y}$)$_{x}$B$_{2}$ where $x=0.8$ to 0.85 and $y\approx 0.4$-0.5, except for one entry with $M=\mrm{Nb}$ at 95\% from Ref.~\cite{Cooper1970}.  
\\
$\dagger$ - midpoint obtained from resistivity drop and not specific heat peak. }
\label{table_2}
\end{table}

\section{Discussion}
Our experiments reveal that rapid cooling synthesis methods, such as the $\rho$-bar and water-cooled splat-quenched techniques, yield higher transition temperatures ($\Tc$) than arc-melted samples. This effect becomes especially pronounced in the case of MoB$_{2}$, which is not 
considered~\cite{Cooper1970} superconducting at ambient pressure. Here, rapid cooling induces a significant superconducting transition at approximately $4.5$ K. Curiously, this matches quite well with the theoretical prediction of 5 K for the $\beta$-phase MoB$_{2}$ at ambient pressure~\cite{Pei2023MoB2}. For alloys like (Zr$_{0.04}$Mo$_{0.96})_{0.85}$B$_{2}$ and Nb$_{0.25}$Mo$_{0.75}$B$_{2}$, the increase in $\Tc$ is less dramatic. However, these samples show some evidence of inhomogeneities, resulting in lower $\Delta C/ (\gamma\Tcmid)$ ratios.

A limitation of our study is the lack of detailed information about the precise structure and composition of the $\rho$-bar and splat-quenched MoB$_{2}$ samples. Understanding the exact B:Mo ratio would provide crucial context for the significance of our findings. Consistent with the literature~\cite{Cooper1970}, we found that MoB$_{2.5}$ is superconducting and is experimentally quite distinct from MoB$_{2}$. The arc-melted and $\rho$-bar MoB$_{2.5}$ samples predominantly show the P$6/mmm$ phase and exhibit broader superconducting transitions. Our results potentially offer the first evidence of ambient-pressure superconductivity in MoB$_{2}$. A careful study of the superconductivity under pressure would be a logical next step. There, we could determine whether our MoB$_{2}$ samples follow an analogous $\D\Tc/\D P$ trend to that discovered by Pei et al.~\cite{Pei2023MoB2} or something else entirely. Whether or not TM-substituted MoB$_{2}$ or other TM-diborides can achieve similar high-$\Tc$ values under lower applied pressure than MoB$_{2}$ remains to be seen. 

While our study primarily investigates the superconducting properties of MoB$_{2}$ and its various alloys, it's important to recognize the broader context. Borides and diborides have long been known to have exemplary high-temperature properties such as high hardness, robust oxidation resistance, and high melting points~\cite{Matkovich1977,Storms1977,Okada1987,Ivanovskii2012,Tao2013,Hayami2013,Xu2014,Ding2016,Tang2020,Magnuson2022}. Recently, they started attracting interest for their low-temperature topological features. A notable example is the emergence of Dirac cones in the electronic structure of monolayer diborides, as highlighted in studies on TiB$_{2}$~\cite{Zhang2014}, FeB$_{2}$~\cite{Zhang2016}, and ZrB$_{2}$~\cite{Lopez-Bezanilla2018}. Adding to this are recently discovered topological features in the phonon spectrum of $\alpha$-MoB$_{2}$, revealing parity-time symmetry-protected helical nodal lines~\cite{Zhang2019}. Such topological states give rise to phononic boundary modes on the surface unaffected by local disorder. However, exploring the potential relevance of this topology of electronic and bosonic states in alloyed TM-diborides, like the ones featured in this work, is still a nascent topic.

\vskip .2cm
\section{Conclusions}  
In this work, we report our experimental results for several TM-substituted MoB$_{2}$ superconductors. As others have shown in previous works, alloying MoB$_{2}$ with other TM's\textemdash especially those with fewer $d$-electrons than Mo\textemdash can help to stabilize the AlB$_{2}$ P6/$mmm$ space group structure at ambient pressure. Seen for the first time, substitutions of 10\% to 50\% Ta and 4\% Hf yield superconducting alloys with $\Tc$'s near 2.4-3.2 K and 6.3 K, respectively. We also examined Zr-substituted MoB$_{2}$ at 4\%, finding $\Tc\sim 7.5$ K similar to older results by Muzzy et al.~\cite{Muzzy2002}. Collectively, the role of TM-substitution into MoB$_{2}$, particularly for elements with fewer $d$-electrons, is to introduce stability by suppressing the antibonding character of dominant Mo-Mo bonds at the Fermi level~\cite{Klesnar1996,Vajeeston2001,Muzzy2002}. One consequence of rapid quenching may be to enhance the electron-phonon coupling $\lambda$, but this increase may occur for different reasons in each material. Rapid cooling appears to lower the Debye temperature in each sample—however, the linear specific heat coefficient trends oppositely in the Nb-doped and excess boron Zr-doped samples. 

The nature and concentration of defects in these samples are almost certainly affected by the speed of the rapid cooling synthesis. Vacancies~\cite{Takeya2004,Shein2006,Ivanovskii2012} and stacking faults~\cite{Muzzy2002,Lim2022} are likely responsible for variations in $\Tc$ in similar TM-diborides. How these defects form under different cooling rates and their role in metastable superconductivity remains an open question.

As expected from the literature, our arc-melted MoB$_{2}$ was not superconducting to 1.7 K. However, two rapidly cooled samples exhibited superconductivity at ambient pressure with $\Tcon\sim 4.5$ K. Although we could not estimate the precise composition and structure from XRD measurements, we showed they are distinct from a known excess boron composition MoB$_{2.5}$. Investigations into the precise composition and properties of rapidly cooled MoB$_{2}$ under high pressure are the subject of future work.

\section*{Acknowledgments}
Work at the University of Florida was performed under the auspices of the U.S. National Science Foundation, Division of Materials Research under Contract No. \ NSF-DMR-2118718.  A.C.H.\ and R.G.H.\ acknowledge support from the Center for Bright Beams, U.S. National Science Foundation award PHY-1549132.).
\textcolor{mag}{}

\setcounter{figure}{0}
\bibliography{ref}

\begin{thebibliography}{40}%
\makeatletter
\providecommand \@ifxundefined [1]{%
 \@ifx{#1\undefined}
}%
\providecommand \@ifnum [1]{%
 \ifnum #1\expandafter \@firstoftwo
 \else \expandafter \@secondoftwo
 \fi
}%
\providecommand \@ifx [1]{%
 \ifx #1\expandafter \@firstoftwo
 \else \expandafter \@secondoftwo
 \fi
}%
\providecommand \natexlab [1]{#1}%
\providecommand \enquote  [1]{``#1''}%
\providecommand \bibnamefont  [1]{#1}%
\providecommand \bibfnamefont [1]{#1}%
\providecommand \citenamefont [1]{#1}%
\providecommand \href@noop [0]{\@secondoftwo}%
\providecommand \href [0]{\begingroup \@sanitize@url \@href}%
\providecommand \@href[1]{\@@startlink{#1}\@@href}%
\providecommand \@@href[1]{\endgroup#1\@@endlink}%
\providecommand \@sanitize@url [0]{\catcode `\\12\catcode `\$12\catcode
  `\&12\catcode `\#12\catcode `\^12\catcode `\_12\catcode `\%12\relax}%
\providecommand \@@startlink[1]{}%
\providecommand \@@endlink[0]{}%
\providecommand \url  [0]{\begingroup\@sanitize@url \@url }%
\providecommand \@url [1]{\endgroup\@href {#1}{\urlprefix }}%
\providecommand \urlprefix  [0]{URL }%
\providecommand \Eprint [0]{\href }%
\providecommand \doibase [0]{https://doi.org/}%
\providecommand \selectlanguage [0]{\@gobble}%
\providecommand \bibinfo  [0]{\@secondoftwo}%
\providecommand \bibfield  [0]{\@secondoftwo}%
\providecommand \translation [1]{[#1]}%
\providecommand \BibitemOpen [0]{}%
\providecommand \bibitemStop [0]{}%
\providecommand \bibitemNoStop [0]{.\EOS\space}%
\providecommand \EOS [0]{\spacefactor3000\relax}%
\providecommand \BibitemShut  [1]{\csname bibitem#1\endcsname}%
\let\auto@bib@innerbib\@empty
\bibitem [{\citenamefont {Nagamatsu}\ \emph {et~al.}(2001)\citenamefont
  {Nagamatsu}, \citenamefont {Nakagawa}, \citenamefont {Muranaka},
  \citenamefont {Zenitani},\ and\ \citenamefont {Akimitsu}}]{Nagamatsu2001}%
  \BibitemOpen
  \bibfield  {author} {\bibinfo {author} {\bibfnamefont {J.}~\bibnamefont
  {Nagamatsu}}, \bibinfo {author} {\bibfnamefont {N.}~\bibnamefont {Nakagawa}},
  \bibinfo {author} {\bibfnamefont {T.}~\bibnamefont {Muranaka}}, \bibinfo
  {author} {\bibfnamefont {Y.}~\bibnamefont {Zenitani}},\ and\ \bibinfo
  {author} {\bibfnamefont {J.}~\bibnamefont {Akimitsu}},\ }\bibfield  {title}
  {\bibinfo {title} {{Superconductivity at 39~K in magnesium diboride}},\
  }\href {https://doi.org/10.1038/35065039} {\bibfield  {journal} {\bibinfo
  {journal} {Nature}\ }\textbf {\bibinfo {volume} {410}},\ \bibinfo {pages}
  {63} (\bibinfo {year} {2001})}\BibitemShut {NoStop}%
\bibitem [{\citenamefont {Tyan}\ \emph {et~al.}(1969)\citenamefont {Tyan},
  \citenamefont {Toth},\ and\ \citenamefont {Chang}}]{Tyan1969}%
  \BibitemOpen
  \bibfield  {author} {\bibinfo {author} {\bibfnamefont {Y.}~\bibnamefont
  {Tyan}}, \bibinfo {author} {\bibfnamefont {L.~E.}\ \bibnamefont {Toth}},\
  and\ \bibinfo {author} {\bibfnamefont {Y.}~\bibnamefont {Chang}},\ }\bibfield
   {title} {\bibinfo {title} {Low temperature specific heat study of the
  electron transfer theory in refractory metal borides},\ }\href
  {https://doi.org/https://doi.org/10.1016/0022-3697(69)90272-8} {\bibfield
  {journal} {\bibinfo  {journal} {Journal of Physics and Chemistry of Solids}\
  }\textbf {\bibinfo {volume} {30}},\ \bibinfo {pages} {785} (\bibinfo {year}
  {1969})}\BibitemShut {NoStop}%
\bibitem [{\citenamefont {Cooper}\ \emph {et~al.}(1970)\citenamefont {Cooper},
  \citenamefont {Corenzwit}, \citenamefont {Longinotti}, \citenamefont
  {Matthias},\ and\ \citenamefont {Zachariasen}}]{Cooper1970}%
  \BibitemOpen
  \bibfield  {author} {\bibinfo {author} {\bibfnamefont {A.~S.}\ \bibnamefont
  {Cooper}}, \bibinfo {author} {\bibfnamefont {E.}~\bibnamefont {Corenzwit}},
  \bibinfo {author} {\bibfnamefont {L.~D.}\ \bibnamefont {Longinotti}},
  \bibinfo {author} {\bibfnamefont {B.~T.}\ \bibnamefont {Matthias}},\ and\
  \bibinfo {author} {\bibfnamefont {W.~H.}\ \bibnamefont {Zachariasen}},\
  }\bibfield  {title} {\bibinfo {title} {Superconductivity: The transition
  temperature peak below four electrons per atom},\ }\href
  {https://doi.org/10.1073/pnas.67.1.313} {\bibfield  {journal} {\bibinfo
  {journal} {Proceedings of the National Academy of Sciences}\ }\textbf
  {\bibinfo {volume} {67}},\ \bibinfo {pages} {313} (\bibinfo {year} {1970})},\
  \Eprint
  {https://arxiv.org/abs/https://www.pnas.org/doi/pdf/10.1073/pnas.67.1.313}
  {https://www.pnas.org/doi/pdf/10.1073/pnas.67.1.313} \BibitemShut {NoStop}%
\bibitem [{\citenamefont {Leyarovska}\ and\ \citenamefont
  {Leyarovski}(1979)}]{Leyarovska1979}%
  \BibitemOpen
  \bibfield  {author} {\bibinfo {author} {\bibfnamefont {L.}~\bibnamefont
  {Leyarovska}}\ and\ \bibinfo {author} {\bibfnamefont {E.}~\bibnamefont
  {Leyarovski}},\ }\bibfield  {title} {\bibinfo {title} {A search for
  superconductivity below 1 k in transition metal borides},\ }\href
  {https://doi.org/https://doi.org/10.1016/0022-5088(79)90100-0} {\bibfield
  {journal} {\bibinfo  {journal} {Journal of the Less Common Metals}\ }\textbf
  {\bibinfo {volume} {67}},\ \bibinfo {pages} {249} (\bibinfo {year}
  {1979})}\BibitemShut {NoStop}%
\bibitem [{\citenamefont {Pei}\ \emph {et~al.}(2023)\citenamefont {Pei},
  \citenamefont {Zhang}, \citenamefont {Wang}, \citenamefont {Zhao},
  \citenamefont {Gao}, \citenamefont {Gong}, \citenamefont {Tian},
  \citenamefont {Luo}, \citenamefont {Li}, \citenamefont {Yang}, \citenamefont
  {Lu}, \citenamefont {Lei}, \citenamefont {Liu},\ and\ \citenamefont
  {Qi}}]{Pei2023MoB2}%
  \BibitemOpen
  \bibfield  {author} {\bibinfo {author} {\bibfnamefont {C.}~\bibnamefont
  {Pei}}, \bibinfo {author} {\bibfnamefont {J.}~\bibnamefont {Zhang}}, \bibinfo
  {author} {\bibfnamefont {Q.}~\bibnamefont {Wang}}, \bibinfo {author}
  {\bibfnamefont {Y.}~\bibnamefont {Zhao}}, \bibinfo {author} {\bibfnamefont
  {L.}~\bibnamefont {Gao}}, \bibinfo {author} {\bibfnamefont {C.}~\bibnamefont
  {Gong}}, \bibinfo {author} {\bibfnamefont {S.}~\bibnamefont {Tian}}, \bibinfo
  {author} {\bibfnamefont {R.}~\bibnamefont {Luo}}, \bibinfo {author}
  {\bibfnamefont {M.}~\bibnamefont {Li}}, \bibinfo {author} {\bibfnamefont
  {W.}~\bibnamefont {Yang}}, \bibinfo {author} {\bibfnamefont {Z.-Y.}\
  \bibnamefont {Lu}}, \bibinfo {author} {\bibfnamefont {H.}~\bibnamefont
  {Lei}}, \bibinfo {author} {\bibfnamefont {K.}~\bibnamefont {Liu}},\ and\
  \bibinfo {author} {\bibfnamefont {Y.}~\bibnamefont {Qi}},\ }\bibfield
  {title} {\bibinfo {title} {{Pressure-induced superconductivity at 32 K in
  MoB$_{2}$}},\ }\href {https://doi.org/10.1093/nsr/nwad034} {\bibfield
  {journal} {\bibinfo  {journal} {National Science Review}\ }\textbf {\bibinfo
  {volume} {10}},\ \bibinfo {pages} {nwad034} (\bibinfo {year} {2023})},\
  \Eprint
  {https://arxiv.org/abs/https://academic.oup.com/nsr/article-pdf/10/5/nwad034/50489440/nwad034.pdf}
  {https://academic.oup.com/nsr/article-pdf/10/5/nwad034/50489440/nwad034.pdf}
  \BibitemShut {NoStop}%
\bibitem [{\citenamefont {Muzzy}\ \emph {et~al.}(2002)\citenamefont {Muzzy},
  \citenamefont {Avdeev}, \citenamefont {Lawes}, \citenamefont {Haas},
  \citenamefont {Zandbergen}, \citenamefont {Ramirez}, \citenamefont
  {Jorgensen},\ and\ \citenamefont {Cava}}]{Muzzy2002}%
  \BibitemOpen
  \bibfield  {author} {\bibinfo {author} {\bibfnamefont {L.}~\bibnamefont
  {Muzzy}}, \bibinfo {author} {\bibfnamefont {M.}~\bibnamefont {Avdeev}},
  \bibinfo {author} {\bibfnamefont {G.}~\bibnamefont {Lawes}}, \bibinfo
  {author} {\bibfnamefont {M.}~\bibnamefont {Haas}}, \bibinfo {author}
  {\bibfnamefont {H.}~\bibnamefont {Zandbergen}}, \bibinfo {author}
  {\bibfnamefont {A.}~\bibnamefont {Ramirez}}, \bibinfo {author} {\bibfnamefont
  {J.}~\bibnamefont {Jorgensen}},\ and\ \bibinfo {author} {\bibfnamefont
  {R.}~\bibnamefont {Cava}},\ }\bibfield  {title} {\bibinfo {title} {Structure
  and superconductivity in {Zr}-stabilized, nonstoichiometric molybdenum
  diboride},\ }\href
  {https://doi.org/https://doi.org/10.1016/S0921-4534(02)01861-0} {\bibfield
  {journal} {\bibinfo  {journal} {Physica C: Superconductivity}\ }\textbf
  {\bibinfo {volume} {382}},\ \bibinfo {pages} {153} (\bibinfo {year}
  {2002})}\BibitemShut {NoStop}%
\bibitem [{\citenamefont {Yang}\ \emph {et~al.}(2022)\citenamefont {Yang},
  \citenamefont {Xiao}, \citenamefont {Zhu}, \citenamefont {Cui}, \citenamefont
  {Song}, \citenamefont {Cao},\ and\ \citenamefont {Ren}}]{Yang2022}%
  \BibitemOpen
  \bibfield  {author} {\bibinfo {author} {\bibfnamefont {W.}~\bibnamefont
  {Yang}}, \bibinfo {author} {\bibfnamefont {G.}~\bibnamefont {Xiao}}, \bibinfo
  {author} {\bibfnamefont {Q.}~\bibnamefont {Zhu}}, \bibinfo {author}
  {\bibfnamefont {Y.}~\bibnamefont {Cui}}, \bibinfo {author} {\bibfnamefont
  {S.}~\bibnamefont {Song}}, \bibinfo {author} {\bibfnamefont {G.-H.}\
  \bibnamefont {Cao}},\ and\ \bibinfo {author} {\bibfnamefont {Z.}~\bibnamefont
  {Ren}},\ }\bibfield  {title} {\bibinfo {title} {{Stabilization and
  superconductivity of AlB$_{2}$-type nonstoichiometric molybdenum diboride by
  Sc doping}},\ }\href
  {https://doi.org/https://doi.org/10.1016/j.ceramint.2022.03.272} {\bibfield
  {journal} {\bibinfo  {journal} {Ceramics International}\ }\textbf {\bibinfo
  {volume} {48}},\ \bibinfo {pages} {19971} (\bibinfo {year}
  {2022})}\BibitemShut {NoStop}%
\bibitem [{\citenamefont {Hire}\ \emph {et~al.}(2022)\citenamefont {Hire},
  \citenamefont {Sinha}, \citenamefont {Lim}, \citenamefont {Kim},
  \citenamefont {Dee}, \citenamefont {Fanfarillo}, \citenamefont {Hamlin},
  \citenamefont {Hennig}, \citenamefont {Hirschfeld},\ and\ \citenamefont
  {Stewart}}]{Hire2022}%
  \BibitemOpen
  \bibfield  {author} {\bibinfo {author} {\bibfnamefont {A.~C.}\ \bibnamefont
  {Hire}}, \bibinfo {author} {\bibfnamefont {S.}~\bibnamefont {Sinha}},
  \bibinfo {author} {\bibfnamefont {J.}~\bibnamefont {Lim}}, \bibinfo {author}
  {\bibfnamefont {J.~S.}\ \bibnamefont {Kim}}, \bibinfo {author} {\bibfnamefont
  {P.~M.}\ \bibnamefont {Dee}}, \bibinfo {author} {\bibfnamefont
  {L.}~\bibnamefont {Fanfarillo}}, \bibinfo {author} {\bibfnamefont {J.~J.}\
  \bibnamefont {Hamlin}}, \bibinfo {author} {\bibfnamefont {R.~G.}\
  \bibnamefont {Hennig}}, \bibinfo {author} {\bibfnamefont {P.~J.}\
  \bibnamefont {Hirschfeld}},\ and\ \bibinfo {author} {\bibfnamefont {G.~R.}\
  \bibnamefont {Stewart}},\ }\bibfield  {title} {\bibinfo {title} {{High
  critical field superconductivity at ambient pressure in ${\mathrm{MoB}}_{2}$
  stabilized in the P6/mmm structure via Nb substitution}},\ }\href
  {https://doi.org/10.1103/PhysRevB.106.174515} {\bibfield  {journal} {\bibinfo
   {journal} {Phys. Rev. B}\ }\textbf {\bibinfo {volume} {106}},\ \bibinfo
  {pages} {174515} (\bibinfo {year} {2022})}\BibitemShut {NoStop}%
\bibitem [{\citenamefont {Lim}\ \emph {et~al.}(2023)\citenamefont {Lim},
  \citenamefont {Sinha}, \citenamefont {Hire}, \citenamefont {Kim},
  \citenamefont {Dee}, \citenamefont {Kumar}, \citenamefont {Popov},
  \citenamefont {Hemley}, \citenamefont {Hennig}, \citenamefont {Hirschfeld},
  \citenamefont {Stewart},\ and\ \citenamefont {Hamlin}}]{Lim2023suppression}%
  \BibitemOpen
  \bibfield  {author} {\bibinfo {author} {\bibfnamefont {J.}~\bibnamefont
  {Lim}}, \bibinfo {author} {\bibfnamefont {S.}~\bibnamefont {Sinha}}, \bibinfo
  {author} {\bibfnamefont {A.~C.}\ \bibnamefont {Hire}}, \bibinfo {author}
  {\bibfnamefont {J.~S.}\ \bibnamefont {Kim}}, \bibinfo {author} {\bibfnamefont
  {P.~M.}\ \bibnamefont {Dee}}, \bibinfo {author} {\bibfnamefont {R.~S.}\
  \bibnamefont {Kumar}}, \bibinfo {author} {\bibfnamefont {D.}~\bibnamefont
  {Popov}}, \bibinfo {author} {\bibfnamefont {R.~J.}\ \bibnamefont {Hemley}},
  \bibinfo {author} {\bibfnamefont {R.~G.}\ \bibnamefont {Hennig}}, \bibinfo
  {author} {\bibfnamefont {P.~J.}\ \bibnamefont {Hirschfeld}}, \bibinfo
  {author} {\bibfnamefont {G.~R.}\ \bibnamefont {Stewart}},\ and\ \bibinfo
  {author} {\bibfnamefont {J.~J.}\ \bibnamefont {Hamlin}},\ }\bibfield  {title}
  {\bibinfo {title} {{Niobium substitution suppresses the superconducting
  critical temperature of pressurized ${\mathrm{MoB}}_{2}$}},\ }\href
  {https://doi.org/10.1103/PhysRevB.108.094501} {\bibfield  {journal} {\bibinfo
   {journal} {Phys. Rev. B}\ }\textbf {\bibinfo {volume} {108}},\ \bibinfo
  {pages} {094501} (\bibinfo {year} {2023})}\BibitemShut {NoStop}%
\bibitem [{\citenamefont {Burdett}\ \emph {et~al.}(1986)\citenamefont
  {Burdett}, \citenamefont {Canadell},\ and\ \citenamefont
  {Miller}}]{Burdett1986}%
  \BibitemOpen
  \bibfield  {author} {\bibinfo {author} {\bibfnamefont {J.~K.}\ \bibnamefont
  {Burdett}}, \bibinfo {author} {\bibfnamefont {E.}~\bibnamefont {Canadell}},\
  and\ \bibinfo {author} {\bibfnamefont {G.~J.}\ \bibnamefont {Miller}},\
  }\bibfield  {title} {\bibinfo {title} {{Electronic structure of
  transition-metal borides with the AlB$_{2}$ structure}},\ }\href
  {https://doi.org/10.1021/ja00281a020} {\bibfield  {journal} {\bibinfo
  {journal} {Journal of the American Chemical Society}\ }\textbf {\bibinfo
  {volume} {108}},\ \bibinfo {pages} {6561} (\bibinfo {year}
  {1986})}\BibitemShut {NoStop}%
\bibitem [{\citenamefont {Stewart}(1983)}]{Stewart1983}%
  \BibitemOpen
  \bibfield  {author} {\bibinfo {author} {\bibfnamefont {G.~R.}\ \bibnamefont
  {Stewart}},\ }\bibfield  {title} {\bibinfo {title} {Measurement of
  low-temperature specific heat},\ }\href {https://doi.org/10.1063/1.1137207}
  {\bibfield  {journal} {\bibinfo  {journal} {Review of Scientific
  Instruments}\ }\textbf {\bibinfo {volume} {54}},\ \bibinfo {pages} {1}
  (\bibinfo {year} {1983})}\BibitemShut {NoStop}%
\bibitem [{\citenamefont {Jain}\ \emph {et~al.}(2013)\citenamefont {Jain},
  \citenamefont {Ong}, \citenamefont {Hautier}, \citenamefont {Chen},
  \citenamefont {Richards}, \citenamefont {Dacek}, \citenamefont {Cholia},
  \citenamefont {Gunter}, \citenamefont {Skinner}, \citenamefont {Ceder},\ and\
  \citenamefont {Persson}}]{MatProj2013}%
  \BibitemOpen
  \bibfield  {author} {\bibinfo {author} {\bibfnamefont {A.}~\bibnamefont
  {Jain}}, \bibinfo {author} {\bibfnamefont {S.~P.}\ \bibnamefont {Ong}},
  \bibinfo {author} {\bibfnamefont {G.}~\bibnamefont {Hautier}}, \bibinfo
  {author} {\bibfnamefont {W.}~\bibnamefont {Chen}}, \bibinfo {author}
  {\bibfnamefont {W.~D.}\ \bibnamefont {Richards}}, \bibinfo {author}
  {\bibfnamefont {S.}~\bibnamefont {Dacek}}, \bibinfo {author} {\bibfnamefont
  {S.}~\bibnamefont {Cholia}}, \bibinfo {author} {\bibfnamefont
  {D.}~\bibnamefont {Gunter}}, \bibinfo {author} {\bibfnamefont
  {D.}~\bibnamefont {Skinner}}, \bibinfo {author} {\bibfnamefont
  {G.}~\bibnamefont {Ceder}},\ and\ \bibinfo {author} {\bibfnamefont {K.~A.}\
  \bibnamefont {Persson}},\ }\bibfield  {title} {\bibinfo {title} {{Commentary:
  The Materials Project: A materials genome approach to accelerating materials
  innovation}},\ }\href {https://doi.org/10.1063/1.4812323} {\bibfield
  {journal} {\bibinfo  {journal} {APL Materials}\ }\textbf {\bibinfo {volume}
  {1}},\ \bibinfo {pages} {011002} (\bibinfo {year} {2013})}\BibitemShut
  {NoStop}%
\bibitem [{\citenamefont {Coleman}(2015)}]{Coleman_2015}%
  \BibitemOpen
  \bibfield  {author} {\bibinfo {author} {\bibfnamefont {P.}~\bibnamefont
  {Coleman}},\ }\href {https://doi.org/10.1017/CBO9781139020916} {\emph
  {\bibinfo {title} {Introduction to Many-Body Physics}}}\ (\bibinfo
  {publisher} {Cambridge University Press},\ \bibinfo {year}
  {2015})\BibitemShut {NoStop}%
\bibitem [{Note1()}]{Note1}%
  \BibitemOpen
  \bibinfo {note} {Formally, $\lambda _{m} := -\protect \mathrm {d}\Sigma
  (\omega )/\protect \mathrm {d}\omega |_{\omega =0}$ where $\Sigma (\omega )$
  is the electron self-energy averaged over the Fermi surface $\Sigma (\omega
  ):=\langle \Sigma (\protect \mathbf {k},\omega ) \rangle _{\protect \mathbf
  {k}\in \protect \mathrm {FS}}$.}\BibitemShut {Stop}%
\bibitem [{\citenamefont {Lin}\ \emph {et~al.}(2011)\citenamefont {Lin},
  \citenamefont {Hsieh}, \citenamefont {Chareev}, \citenamefont {Vasiliev},
  \citenamefont {Parsons},\ and\ \citenamefont {Yang}}]{Lin2011}%
  \BibitemOpen
  \bibfield  {author} {\bibinfo {author} {\bibfnamefont {J.-Y.}\ \bibnamefont
  {Lin}}, \bibinfo {author} {\bibfnamefont {Y.~S.}\ \bibnamefont {Hsieh}},
  \bibinfo {author} {\bibfnamefont {D.~A.}\ \bibnamefont {Chareev}}, \bibinfo
  {author} {\bibfnamefont {A.~N.}\ \bibnamefont {Vasiliev}}, \bibinfo {author}
  {\bibfnamefont {Y.}~\bibnamefont {Parsons}},\ and\ \bibinfo {author}
  {\bibfnamefont {H.~D.}\ \bibnamefont {Yang}},\ }\bibfield  {title} {\bibinfo
  {title} {{Coexistence of isotropic and extended $s$-wave order parameters in
  FeSe as revealed by low-temperature specific heat}},\ }\href
  {https://doi.org/10.1103/PhysRevB.84.220507} {\bibfield  {journal} {\bibinfo
  {journal} {Phys. Rev. B}\ }\textbf {\bibinfo {volume} {84}},\ \bibinfo
  {pages} {220507} (\bibinfo {year} {2011})}\BibitemShut {NoStop}%
\bibitem [{\citenamefont {McMillan}(1968)}]{McMillan1968}%
  \BibitemOpen
  \bibfield  {author} {\bibinfo {author} {\bibfnamefont {W.~L.}\ \bibnamefont
  {McMillan}},\ }\bibfield  {title} {\bibinfo {title} {Transition temperature
  of strong-coupled superconductors},\ }\href
  {https://doi.org/10.1103/PhysRev.167.331} {\bibfield  {journal} {\bibinfo
  {journal} {Phys. Rev.}\ }\textbf {\bibinfo {volume} {167}},\ \bibinfo {pages}
  {331} (\bibinfo {year} {1968})}\BibitemShut {NoStop}%
\bibitem [{\citenamefont {Quan}\ \emph {et~al.}(2021)\citenamefont {Quan},
  \citenamefont {Lee},\ and\ \citenamefont {Pickett}}]{Quan2021}%
  \BibitemOpen
  \bibfield  {author} {\bibinfo {author} {\bibfnamefont {Y.}~\bibnamefont
  {Quan}}, \bibinfo {author} {\bibfnamefont {K.-W.}\ \bibnamefont {Lee}},\ and\
  \bibinfo {author} {\bibfnamefont {W.~E.}\ \bibnamefont {Pickett}},\
  }\bibfield  {title} {\bibinfo {title} {{MoB$_2$ under pressure:
  Superconducting Mo enhanced by boron}},\ }\href
  {https://doi.org/10.1103/PhysRevB.104.224504} {\bibfield  {journal} {\bibinfo
   {journal} {Phys. Rev. B}\ }\textbf {\bibinfo {volume} {104}},\ \bibinfo
  {pages} {224504} (\bibinfo {year} {2021})}\BibitemShut {NoStop}%
\bibitem [{\citenamefont {Shein}\ and\ \citenamefont
  {Ivanovskii}(2006)}]{Shein2006}%
  \BibitemOpen
  \bibfield  {author} {\bibinfo {author} {\bibfnamefont {I.~R.}\ \bibnamefont
  {Shein}}\ and\ \bibinfo {author} {\bibfnamefont {A.~L.}\ \bibnamefont
  {Ivanovskii}},\ }\bibfield  {title} {\bibinfo {title} {Influence of lattice
  vacancies on the structural, electronic, and cohesive properties of niobium
  and molybdenum borides from first-principles calculations},\ }\href
  {https://doi.org/10.1103/PhysRevB.73.144108} {\bibfield  {journal} {\bibinfo
  {journal} {Phys. Rev. B}\ }\textbf {\bibinfo {volume} {73}},\ \bibinfo
  {pages} {144108} (\bibinfo {year} {2006})}\BibitemShut {NoStop}%
\bibitem [{\citenamefont {Heid}\ \emph {et~al.}(2003)\citenamefont {Heid},
  \citenamefont {Renker}, \citenamefont {Schober}, \citenamefont {Adelmann},
  \citenamefont {Ernst},\ and\ \citenamefont {Bohnen}}]{Heid2003}%
  \BibitemOpen
  \bibfield  {author} {\bibinfo {author} {\bibfnamefont {R.}~\bibnamefont
  {Heid}}, \bibinfo {author} {\bibfnamefont {B.}~\bibnamefont {Renker}},
  \bibinfo {author} {\bibfnamefont {H.}~\bibnamefont {Schober}}, \bibinfo
  {author} {\bibfnamefont {P.}~\bibnamefont {Adelmann}}, \bibinfo {author}
  {\bibfnamefont {D.}~\bibnamefont {Ernst}},\ and\ \bibinfo {author}
  {\bibfnamefont {K.-P.}\ \bibnamefont {Bohnen}},\ }\bibfield  {title}
  {\bibinfo {title} {Lattice dynamics and electron-phonon coupling in
  transition-metal diborides},\ }\href
  {https://doi.org/10.1103/PhysRevB.67.180510} {\bibfield  {journal} {\bibinfo
  {journal} {Phys. Rev. B}\ }\textbf {\bibinfo {volume} {67}},\ \bibinfo
  {pages} {180510} (\bibinfo {year} {2003})}\BibitemShut {NoStop}%
\bibitem [{\citenamefont {Allen}\ and\ \citenamefont
  {Dynes}(1975)}]{Allen-Dynes1975}%
  \BibitemOpen
  \bibfield  {author} {\bibinfo {author} {\bibfnamefont {P.~B.}\ \bibnamefont
  {Allen}}\ and\ \bibinfo {author} {\bibfnamefont {R.~C.}\ \bibnamefont
  {Dynes}},\ }\bibfield  {title} {\bibinfo {title} {Transition temperature of
  strong-coupled superconductors reanalyzed},\ }\href
  {https://doi.org/10.1103/PhysRevB.12.905} {\bibfield  {journal} {\bibinfo
  {journal} {Phys. Rev. B}\ }\textbf {\bibinfo {volume} {12}},\ \bibinfo
  {pages} {905} (\bibinfo {year} {1975})}\BibitemShut {NoStop}%
\bibitem [{\citenamefont {Lim}\ \emph {et~al.}(2022)\citenamefont {Lim},
  \citenamefont {Hire}, \citenamefont {Quan}, \citenamefont {Kim},
  \citenamefont {Xie}, \citenamefont {Sinha}, \citenamefont {Kumar},
  \citenamefont {Popov}, \citenamefont {Park}, \citenamefont {Hemley},
  \citenamefont {Vohra}, \citenamefont {Hamlin}, \citenamefont {Hennig},
  \citenamefont {Hirschfeld},\ and\ \citenamefont {Stewart}}]{Lim2022}%
  \BibitemOpen
  \bibfield  {author} {\bibinfo {author} {\bibfnamefont {J.}~\bibnamefont
  {Lim}}, \bibinfo {author} {\bibfnamefont {A.~C.}\ \bibnamefont {Hire}},
  \bibinfo {author} {\bibfnamefont {Y.}~\bibnamefont {Quan}}, \bibinfo {author}
  {\bibfnamefont {J.~S.}\ \bibnamefont {Kim}}, \bibinfo {author} {\bibfnamefont
  {S.~R.}\ \bibnamefont {Xie}}, \bibinfo {author} {\bibfnamefont
  {S.}~\bibnamefont {Sinha}}, \bibinfo {author} {\bibfnamefont {R.~S.}\
  \bibnamefont {Kumar}}, \bibinfo {author} {\bibfnamefont {D.}~\bibnamefont
  {Popov}}, \bibinfo {author} {\bibfnamefont {C.}~\bibnamefont {Park}},
  \bibinfo {author} {\bibfnamefont {R.~J.}\ \bibnamefont {Hemley}}, \bibinfo
  {author} {\bibfnamefont {Y.~K.}\ \bibnamefont {Vohra}}, \bibinfo {author}
  {\bibfnamefont {J.~J.}\ \bibnamefont {Hamlin}}, \bibinfo {author}
  {\bibfnamefont {R.~G.}\ \bibnamefont {Hennig}}, \bibinfo {author}
  {\bibfnamefont {P.~J.}\ \bibnamefont {Hirschfeld}},\ and\ \bibinfo {author}
  {\bibfnamefont {G.~R.}\ \bibnamefont {Stewart}},\ }\bibfield  {title}
  {\bibinfo {title} {Creating superconductivity in wb2 through pressure-induced
  metastable planar defects},\ }\href
  {https://doi.org/10.1038/s41467-022-35191-8} {\bibfield  {journal} {\bibinfo
  {journal} {Nature Communications}\ }\textbf {\bibinfo {volume} {13}},\
  \bibinfo {pages} {7901} (\bibinfo {year} {2022})}\BibitemShut {NoStop}%
\bibitem [{\citenamefont {Liu}\ \emph {et~al.}(2022)\citenamefont {Liu},
  \citenamefont {Huang}, \citenamefont {Song}, \citenamefont {Wang},
  \citenamefont {Zhang}, \citenamefont {Lv}, \citenamefont {Liu}, \citenamefont
  {Zhang}, \citenamefont {Cho},\ and\ \citenamefont {Jia}}]{Liu2022}%
  \BibitemOpen
  \bibfield  {author} {\bibinfo {author} {\bibfnamefont {X.}~\bibnamefont
  {Liu}}, \bibinfo {author} {\bibfnamefont {X.}~\bibnamefont {Huang}}, \bibinfo
  {author} {\bibfnamefont {P.}~\bibnamefont {Song}}, \bibinfo {author}
  {\bibfnamefont {C.}~\bibnamefont {Wang}}, \bibinfo {author} {\bibfnamefont
  {L.}~\bibnamefont {Zhang}}, \bibinfo {author} {\bibfnamefont
  {P.}~\bibnamefont {Lv}}, \bibinfo {author} {\bibfnamefont {L.}~\bibnamefont
  {Liu}}, \bibinfo {author} {\bibfnamefont {W.}~\bibnamefont {Zhang}}, \bibinfo
  {author} {\bibfnamefont {J.-H.}\ \bibnamefont {Cho}},\ and\ \bibinfo {author}
  {\bibfnamefont {Y.}~\bibnamefont {Jia}},\ }\bibfield  {title} {\bibinfo
  {title} {{Strong electron-phonon coupling superconductivity in compressed
  $\ensuremath{\alpha}\text{\ensuremath{-}}{\mathrm{MoB}}_{2}$ induced by
  double Van Hove singularities}},\ }\href
  {https://doi.org/10.1103/PhysRevB.106.064507} {\bibfield  {journal} {\bibinfo
   {journal} {Phys. Rev. B}\ }\textbf {\bibinfo {volume} {106}},\ \bibinfo
  {pages} {064507} (\bibinfo {year} {2022})}\BibitemShut {NoStop}%
\bibitem [{\citenamefont {Klesnar}\ \emph {et~al.}(1996)\citenamefont
  {Klesnar}, \citenamefont {Aselage}, \citenamefont {Morosin},\ and\
  \citenamefont {Kwei}}]{Klesnar1996}%
  \BibitemOpen
  \bibfield  {author} {\bibinfo {author} {\bibfnamefont {H.}~\bibnamefont
  {Klesnar}}, \bibinfo {author} {\bibfnamefont {T.}~\bibnamefont {Aselage}},
  \bibinfo {author} {\bibfnamefont {B.}~\bibnamefont {Morosin}},\ and\ \bibinfo
  {author} {\bibfnamefont {G.}~\bibnamefont {Kwei}},\ }\bibfield  {title}
  {\bibinfo {title} {{The diboride compounds of molybdenum: MoB$_{2-x}$ and
  Mo$_{2}$B$_{5-y}$}},\ }\href
  {https://doi.org/https://doi.org/10.1016/0925-8388(96)02294-3} {\bibfield
  {journal} {\bibinfo  {journal} {Journal of Alloys and Compounds}\ }\textbf
  {\bibinfo {volume} {241}},\ \bibinfo {pages} {180} (\bibinfo {year}
  {1996})}\BibitemShut {NoStop}%
\bibitem [{\citenamefont {Yang}\ \emph {et~al.}(2023)\citenamefont {Yang},
  \citenamefont {Xiao}, \citenamefont {Zhu}, \citenamefont {Song},
  \citenamefont {Cao},\ and\ \citenamefont {Ren}}]{Yang2023}%
  \BibitemOpen
  \bibfield  {author} {\bibinfo {author} {\bibfnamefont {W.}~\bibnamefont
  {Yang}}, \bibinfo {author} {\bibfnamefont {G.}~\bibnamefont {Xiao}}, \bibinfo
  {author} {\bibfnamefont {Q.}~\bibnamefont {Zhu}}, \bibinfo {author}
  {\bibfnamefont {S.}~\bibnamefont {Song}}, \bibinfo {author} {\bibfnamefont
  {G.-H.}\ \bibnamefont {Cao}},\ and\ \bibinfo {author} {\bibfnamefont
  {Z.}~\bibnamefont {Ren}},\ }\bibfield  {title} {\bibinfo {title} {{Effect of
  carbon doping on the structure and superconductivity in AlB$_{2}$-type
  (Mo$_{0.96}$Ti$_{0.04}$)$_{0.8}$B$_{2}$}},\ }\href
  {https://doi.org/https://doi.org/10.1111/jace.19062} {\bibfield  {journal}
  {\bibinfo  {journal} {Journal of the American Ceramic Society}\ }\textbf
  {\bibinfo {volume} {106}},\ \bibinfo {pages} {4211} (\bibinfo {year}
  {2023})}\BibitemShut {NoStop}%
\bibitem [{\citenamefont {Matkovich}(1977)}]{Matkovich1977}%
  \BibitemOpen
  \bibinfo {editor} {\bibfnamefont {V.~I.}\ \bibnamefont {Matkovich}},\ ed.,\
  \href {https://doi.org/https://doi.org/10.1007/978-3-642-66620-9} {\emph
  {\bibinfo {title} {Boron and Refractory Borides}}},\ \bibinfo {edition}
  {1st}\ ed.\ (\bibinfo  {publisher} {Springer Berlin, Heidelberg},\ \bibinfo
  {year} {1977})\ pp.\ \bibinfo {pages} {X, 656},\ \bibinfo {note} {copyright:
  Springer-Verlag Berlin Heidelberg 1977}\BibitemShut {NoStop}%
\bibitem [{\citenamefont {Storms}\ and\ \citenamefont
  {Mueller}(1977)}]{Storms1977}%
  \BibitemOpen
  \bibfield  {author} {\bibinfo {author} {\bibfnamefont {E.}~\bibnamefont
  {Storms}}\ and\ \bibinfo {author} {\bibfnamefont {B.}~\bibnamefont
  {Mueller}},\ }\bibfield  {title} {\bibinfo {title} {{Phase relations and
  thermodynamic properties of transition metal borides. I. The molybdenum-boron
  system and elemental boron.}},\ }\href {https://doi.org/10.1021/j100519a008}
  {\bibfield  {journal} {\bibinfo  {journal} {The Journal of Physical
  Chemistry}\ }\textbf {\bibinfo {volume} {81}},\ \bibinfo {pages} {318}
  (\bibinfo {year} {1977})}\BibitemShut {NoStop}%
\bibitem [{\citenamefont {Okada}\ \emph {et~al.}(1987)\citenamefont {Okada},
  \citenamefont {Atoda}, \citenamefont {Higashi},\ and\ \citenamefont
  {Takahashi}}]{Okada1987}%
  \BibitemOpen
  \bibfield  {author} {\bibinfo {author} {\bibfnamefont {S.}~\bibnamefont
  {Okada}}, \bibinfo {author} {\bibfnamefont {T.}~\bibnamefont {Atoda}},
  \bibinfo {author} {\bibfnamefont {I.}~\bibnamefont {Higashi}},\ and\ \bibinfo
  {author} {\bibfnamefont {Y.}~\bibnamefont {Takahashi}},\ }\bibfield  {title}
  {\bibinfo {title} {Preparation of single crystals of mob2 by the
  aluminium-flux technique and some of their properties},\ }\href
  {https://doi.org/10.1007/BF01086503} {\bibfield  {journal} {\bibinfo
  {journal} {Journal of Materials Science}\ }\textbf {\bibinfo {volume} {22}},\
  \bibinfo {pages} {2993} (\bibinfo {year} {1987})}\BibitemShut {NoStop}%
\bibitem [{\citenamefont {Ivanovskii}(2012)}]{Ivanovskii2012}%
  \BibitemOpen
  \bibfield  {author} {\bibinfo {author} {\bibfnamefont {A.}~\bibnamefont
  {Ivanovskii}},\ }\bibfield  {title} {\bibinfo {title} {Mechanical and
  electronic properties of diborides of transition 3d–5d metals from first
  principles: Toward search of novel ultra-incompressible and superhard
  materials},\ }\href
  {https://doi.org/https://doi.org/10.1016/j.pmatsci.2011.05.004} {\bibfield
  {journal} {\bibinfo  {journal} {Progress in Materials Science}\ }\textbf
  {\bibinfo {volume} {57}},\ \bibinfo {pages} {184} (\bibinfo {year}
  {2012})}\BibitemShut {NoStop}%
\bibitem [{\citenamefont {Tao}\ \emph {et~al.}(2013)\citenamefont {Tao},
  \citenamefont {Zhao}, \citenamefont {Chen}, \citenamefont {Li}, \citenamefont
  {Li}, \citenamefont {Ma}, \citenamefont {Li}, \citenamefont {Cui},
  \citenamefont {Zhu},\ and\ \citenamefont {Wang}}]{Tao2013}%
  \BibitemOpen
  \bibfield  {author} {\bibinfo {author} {\bibfnamefont {Q.}~\bibnamefont
  {Tao}}, \bibinfo {author} {\bibfnamefont {X.}~\bibnamefont {Zhao}}, \bibinfo
  {author} {\bibfnamefont {Y.}~\bibnamefont {Chen}}, \bibinfo {author}
  {\bibfnamefont {J.}~\bibnamefont {Li}}, \bibinfo {author} {\bibfnamefont
  {Q.}~\bibnamefont {Li}}, \bibinfo {author} {\bibfnamefont {Y.}~\bibnamefont
  {Ma}}, \bibinfo {author} {\bibfnamefont {J.}~\bibnamefont {Li}}, \bibinfo
  {author} {\bibfnamefont {T.}~\bibnamefont {Cui}}, \bibinfo {author}
  {\bibfnamefont {P.}~\bibnamefont {Zhu}},\ and\ \bibinfo {author}
  {\bibfnamefont {X.}~\bibnamefont {Wang}},\ }\bibfield  {title} {\bibinfo
  {title} {Enhanced vickers hardness by quasi-3d boron network in mob2},\
  }\href {https://doi.org/10.1039/C3RA42741B} {\bibfield  {journal} {\bibinfo
  {journal} {RSC Adv.}\ }\textbf {\bibinfo {volume} {3}},\ \bibinfo {pages}
  {18317} (\bibinfo {year} {2013})}\BibitemShut {NoStop}%
\bibitem [{\citenamefont {Hayami}\ \emph {et~al.}(2013)\citenamefont {Hayami},
  \citenamefont {Momozawa},\ and\ \citenamefont {Otani}}]{Hayami2013}%
  \BibitemOpen
  \bibfield  {author} {\bibinfo {author} {\bibfnamefont {W.}~\bibnamefont
  {Hayami}}, \bibinfo {author} {\bibfnamefont {A.}~\bibnamefont {Momozawa}},\
  and\ \bibinfo {author} {\bibfnamefont {S.}~\bibnamefont {Otani}},\ }\bibfield
   {title} {\bibinfo {title} {{Effect of Defects in the Formation of
  AlB$_{2}$-Type WB$_{2}$ and MoB$_{2}$}},\ }\href
  {https://doi.org/10.1021/ic400587j} {\bibfield  {journal} {\bibinfo
  {journal} {Inorganic Chemistry}\ }\textbf {\bibinfo {volume} {52}},\ \bibinfo
  {pages} {7573} (\bibinfo {year} {2013})},\ \bibinfo {note} {pMID:
  24004287}\BibitemShut {NoStop}%
\bibitem [{\citenamefont {Xu}\ \emph {et~al.}(2014)\citenamefont {Xu},
  \citenamefont {Fu}, \citenamefont {Yu}, \citenamefont {Lu}, \citenamefont
  {Zhang}, \citenamefont {Liu},\ and\ \citenamefont {Tang}}]{Xu2014}%
  \BibitemOpen
  \bibfield  {author} {\bibinfo {author} {\bibfnamefont {X.}~\bibnamefont
  {Xu}}, \bibinfo {author} {\bibfnamefont {K.}~\bibnamefont {Fu}}, \bibinfo
  {author} {\bibfnamefont {M.}~\bibnamefont {Yu}}, \bibinfo {author}
  {\bibfnamefont {Z.}~\bibnamefont {Lu}}, \bibinfo {author} {\bibfnamefont
  {X.}~\bibnamefont {Zhang}}, \bibinfo {author} {\bibfnamefont
  {G.}~\bibnamefont {Liu}},\ and\ \bibinfo {author} {\bibfnamefont
  {C.}~\bibnamefont {Tang}},\ }\bibfield  {title} {\bibinfo {title} {The
  thermodynamic, electronic and elastic properties of the
  early-transition-metal diborides with alb2-type structure: A density
  functional theory study},\ }\href
  {https://doi.org/https://doi.org/10.1016/j.jallcom.2014.04.067} {\bibfield
  {journal} {\bibinfo  {journal} {Journal of Alloys and Compounds}\ }\textbf
  {\bibinfo {volume} {607}},\ \bibinfo {pages} {198} (\bibinfo {year}
  {2014})}\BibitemShut {NoStop}%
\bibitem [{\citenamefont {Ding}\ \emph {et~al.}(2016)\citenamefont {Ding},
  \citenamefont {Shao}, \citenamefont {Zhang}, \citenamefont {Lu},
  \citenamefont {Ding}, \citenamefont {Ning},\ and\ \citenamefont
  {Huang}}]{Ding2016}%
  \BibitemOpen
  \bibfield  {author} {\bibinfo {author} {\bibfnamefont {L.-P.}\ \bibnamefont
  {Ding}}, \bibinfo {author} {\bibfnamefont {P.}~\bibnamefont {Shao}}, \bibinfo
  {author} {\bibfnamefont {F.-H.}\ \bibnamefont {Zhang}}, \bibinfo {author}
  {\bibfnamefont {C.}~\bibnamefont {Lu}}, \bibinfo {author} {\bibfnamefont
  {L.}~\bibnamefont {Ding}}, \bibinfo {author} {\bibfnamefont {S.~Y.}\
  \bibnamefont {Ning}},\ and\ \bibinfo {author} {\bibfnamefont {X.~F.}\
  \bibnamefont {Huang}},\ }\bibfield  {title} {\bibinfo {title} {Crystal
  structures, stabilities, electronic properties, and hardness of mob2:
  First-principles calculations},\ }\href
  {https://doi.org/10.1021/acs.inorgchem.6b00899} {\bibfield  {journal}
  {\bibinfo  {journal} {Inorganic Chemistry}\ }\textbf {\bibinfo {volume}
  {55}},\ \bibinfo {pages} {7033} (\bibinfo {year} {2016})},\ \bibinfo {note}
  {pMID: 27387577},\ \Eprint
  {https://arxiv.org/abs/https://doi.org/10.1021/acs.inorgchem.6b00899}
  {https://doi.org/10.1021/acs.inorgchem.6b00899} \BibitemShut {NoStop}%
\bibitem [{\citenamefont {Tang}\ \emph {et~al.}(2020)\citenamefont {Tang},
  \citenamefont {Gao}, \citenamefont {Zhang}, \citenamefont {Gao},
  \citenamefont {Zhou}, \citenamefont {Yan}, \citenamefont {Li}, \citenamefont
  {Zhang}, \citenamefont {Peng}, \citenamefont {Huang}, \citenamefont {Zhang},
  \citenamefont {Yuan}, \citenamefont {Wan}, \citenamefont {Peng},
  \citenamefont {Wu}, \citenamefont {Zhang}, \citenamefont {Liu}, \citenamefont
  {Gu}, \citenamefont {Gao}, \citenamefont {Irifune}, \citenamefont {Ahuja},
  \citenamefont {Mao},\ and\ \citenamefont {Gou}}]{Tang2020}%
  \BibitemOpen
  \bibfield  {author} {\bibinfo {author} {\bibfnamefont {H.}~\bibnamefont
  {Tang}}, \bibinfo {author} {\bibfnamefont {X.}~\bibnamefont {Gao}}, \bibinfo
  {author} {\bibfnamefont {J.}~\bibnamefont {Zhang}}, \bibinfo {author}
  {\bibfnamefont {B.}~\bibnamefont {Gao}}, \bibinfo {author} {\bibfnamefont
  {W.}~\bibnamefont {Zhou}}, \bibinfo {author} {\bibfnamefont {B.}~\bibnamefont
  {Yan}}, \bibinfo {author} {\bibfnamefont {X.}~\bibnamefont {Li}}, \bibinfo
  {author} {\bibfnamefont {Q.}~\bibnamefont {Zhang}}, \bibinfo {author}
  {\bibfnamefont {S.}~\bibnamefont {Peng}}, \bibinfo {author} {\bibfnamefont
  {D.}~\bibnamefont {Huang}}, \bibinfo {author} {\bibfnamefont
  {L.}~\bibnamefont {Zhang}}, \bibinfo {author} {\bibfnamefont
  {X.}~\bibnamefont {Yuan}}, \bibinfo {author} {\bibfnamefont {B.}~\bibnamefont
  {Wan}}, \bibinfo {author} {\bibfnamefont {C.}~\bibnamefont {Peng}}, \bibinfo
  {author} {\bibfnamefont {L.}~\bibnamefont {Wu}}, \bibinfo {author}
  {\bibfnamefont {D.}~\bibnamefont {Zhang}}, \bibinfo {author} {\bibfnamefont
  {H.}~\bibnamefont {Liu}}, \bibinfo {author} {\bibfnamefont {L.}~\bibnamefont
  {Gu}}, \bibinfo {author} {\bibfnamefont {F.}~\bibnamefont {Gao}}, \bibinfo
  {author} {\bibfnamefont {T.}~\bibnamefont {Irifune}}, \bibinfo {author}
  {\bibfnamefont {R.}~\bibnamefont {Ahuja}}, \bibinfo {author} {\bibfnamefont
  {H.-K.}\ \bibnamefont {Mao}},\ and\ \bibinfo {author} {\bibfnamefont
  {H.}~\bibnamefont {Gou}},\ }\bibfield  {title} {\bibinfo {title} {Boron-rich
  molybdenum boride with unusual short-range vacancy ordering, anisotropic
  hardness, and superconductivity},\ }\href
  {https://doi.org/10.1021/acs.chemmater.9b04052} {\bibfield  {journal}
  {\bibinfo  {journal} {Chemistry of Materials}\ }\textbf {\bibinfo {volume}
  {32}},\ \bibinfo {pages} {459} (\bibinfo {year} {2020})},\ \Eprint
  {https://arxiv.org/abs/https://doi.org/10.1021/acs.chemmater.9b04052}
  {https://doi.org/10.1021/acs.chemmater.9b04052} \BibitemShut {NoStop}%
\bibitem [{\citenamefont {Magnuson}\ \emph {et~al.}(2022)\citenamefont
  {Magnuson}, \citenamefont {Hultman},\ and\ \citenamefont
  {Högberg}}]{Magnuson2022}%
  \BibitemOpen
  \bibfield  {author} {\bibinfo {author} {\bibfnamefont {M.}~\bibnamefont
  {Magnuson}}, \bibinfo {author} {\bibfnamefont {L.}~\bibnamefont {Hultman}},\
  and\ \bibinfo {author} {\bibfnamefont {H.}~\bibnamefont {Högberg}},\
  }\bibfield  {title} {\bibinfo {title} {Review of transition-metal diboride
  thin films},\ }\href
  {https://doi.org/https://doi.org/10.1016/j.vacuum.2021.110567} {\bibfield
  {journal} {\bibinfo  {journal} {Vacuum}\ }\textbf {\bibinfo {volume} {196}},\
  \bibinfo {pages} {110567} (\bibinfo {year} {2022})}\BibitemShut {NoStop}%
\bibitem [{\citenamefont {Zhang}\ \emph {et~al.}(2014)\citenamefont {Zhang},
  \citenamefont {Wang}, \citenamefont {Du}, \citenamefont {Gao},\ and\
  \citenamefont {Liu}}]{Zhang2014}%
  \BibitemOpen
  \bibfield  {author} {\bibinfo {author} {\bibfnamefont {L.~Z.}\ \bibnamefont
  {Zhang}}, \bibinfo {author} {\bibfnamefont {Z.~F.}\ \bibnamefont {Wang}},
  \bibinfo {author} {\bibfnamefont {S.~X.}\ \bibnamefont {Du}}, \bibinfo
  {author} {\bibfnamefont {H.-J.}\ \bibnamefont {Gao}},\ and\ \bibinfo {author}
  {\bibfnamefont {F.}~\bibnamefont {Liu}},\ }\bibfield  {title} {\bibinfo
  {title} {{Prediction of a Dirac state in monolayer ${\mathrm{TiB}}_{2}$}},\
  }\href {https://doi.org/10.1103/PhysRevB.90.161402} {\bibfield  {journal}
  {\bibinfo  {journal} {Phys. Rev. B}\ }\textbf {\bibinfo {volume} {90}},\
  \bibinfo {pages} {161402} (\bibinfo {year} {2014})}\BibitemShut {NoStop}%
\bibitem [{\citenamefont {Zhang}\ \emph {et~al.}(2016)\citenamefont {Zhang},
  \citenamefont {Li}, \citenamefont {Hou}, \citenamefont {Du},\ and\
  \citenamefont {Chen}}]{Zhang2016}%
  \BibitemOpen
  \bibfield  {author} {\bibinfo {author} {\bibfnamefont {H.}~\bibnamefont
  {Zhang}}, \bibinfo {author} {\bibfnamefont {Y.}~\bibnamefont {Li}}, \bibinfo
  {author} {\bibfnamefont {J.}~\bibnamefont {Hou}}, \bibinfo {author}
  {\bibfnamefont {A.}~\bibnamefont {Du}},\ and\ \bibinfo {author}
  {\bibfnamefont {Z.}~\bibnamefont {Chen}},\ }\bibfield  {title} {\bibinfo
  {title} {{Dirac State in the FeB2 Monolayer with Graphene-Like Boron
  Sheet}},\ }\href {https://doi.org/10.1021/acs.nanolett.6b02335} {\bibfield
  {journal} {\bibinfo  {journal} {Nano Letters}\ }\textbf {\bibinfo {volume}
  {16}},\ \bibinfo {pages} {6124} (\bibinfo {year} {2016})}\BibitemShut
  {NoStop}%
\bibitem [{\citenamefont {Lopez-Bezanilla}(2018)}]{Lopez-Bezanilla2018}%
  \BibitemOpen
  \bibfield  {author} {\bibinfo {author} {\bibfnamefont {A.}~\bibnamefont
  {Lopez-Bezanilla}},\ }\bibfield  {title} {\bibinfo {title} {{Twelve
  inequivalent Dirac cones in two-dimensional ${\mathrm{ZrB}}_{2}$}},\ }\href
  {https://doi.org/10.1103/PhysRevMaterials.2.011002} {\bibfield  {journal}
  {\bibinfo  {journal} {Phys. Rev. Mater.}\ }\textbf {\bibinfo {volume} {2}},\
  \bibinfo {pages} {011002} (\bibinfo {year} {2018})}\BibitemShut {NoStop}%
\bibitem [{\citenamefont {Zhang}\ \emph {et~al.}(2019)\citenamefont {Zhang},
  \citenamefont {Miao}, \citenamefont {Wang}, \citenamefont {Lin},
  \citenamefont {Cao}, \citenamefont {Fabbris}, \citenamefont {Said},
  \citenamefont {Liu}, \citenamefont {Lei}, \citenamefont {Fang}, \citenamefont
  {Weng},\ and\ \citenamefont {Dean}}]{Zhang2019}%
  \BibitemOpen
  \bibfield  {author} {\bibinfo {author} {\bibfnamefont {T.~T.}\ \bibnamefont
  {Zhang}}, \bibinfo {author} {\bibfnamefont {H.}~\bibnamefont {Miao}},
  \bibinfo {author} {\bibfnamefont {Q.}~\bibnamefont {Wang}}, \bibinfo {author}
  {\bibfnamefont {J.~Q.}\ \bibnamefont {Lin}}, \bibinfo {author} {\bibfnamefont
  {Y.}~\bibnamefont {Cao}}, \bibinfo {author} {\bibfnamefont {G.}~\bibnamefont
  {Fabbris}}, \bibinfo {author} {\bibfnamefont {A.~H.}\ \bibnamefont {Said}},
  \bibinfo {author} {\bibfnamefont {X.}~\bibnamefont {Liu}}, \bibinfo {author}
  {\bibfnamefont {H.~C.}\ \bibnamefont {Lei}}, \bibinfo {author} {\bibfnamefont
  {Z.}~\bibnamefont {Fang}}, \bibinfo {author} {\bibfnamefont {H.~M.}\
  \bibnamefont {Weng}},\ and\ \bibinfo {author} {\bibfnamefont {M.~P.~M.}\
  \bibnamefont {Dean}},\ }\bibfield  {title} {\bibinfo {title} {{Phononic
  Helical Nodal Lines with $\mathcal{PT}$ Protection in
  ${\mathrm{MoB}}_{2}$}},\ }\href
  {https://doi.org/10.1103/PhysRevLett.123.245302} {\bibfield  {journal}
  {\bibinfo  {journal} {Phys. Rev. Lett.}\ }\textbf {\bibinfo {volume} {123}},\
  \bibinfo {pages} {245302} (\bibinfo {year} {2019})}\BibitemShut {NoStop}%
\bibitem [{\citenamefont {Vajeeston}\ \emph {et~al.}(2001)\citenamefont
  {Vajeeston}, \citenamefont {Ravindran}, \citenamefont {Ravi},\ and\
  \citenamefont {Asokamani1}}]{Vajeeston2001}%
  \BibitemOpen
  \bibfield  {author} {\bibinfo {author} {\bibfnamefont {P.}~\bibnamefont
  {Vajeeston}}, \bibinfo {author} {\bibfnamefont {P.}~\bibnamefont
  {Ravindran}}, \bibinfo {author} {\bibfnamefont {C.}~\bibnamefont {Ravi}},\
  and\ \bibinfo {author} {\bibfnamefont {R.}~\bibnamefont {Asokamani1}},\
  }\bibfield  {title} {\bibinfo {title} {Electronic structure, bonding, and
  ground-state properties of ${\mathrm{alb}}_{2}$-type transition-metal
  diborides},\ }\href {https://doi.org/10.1103/PhysRevB.63.045115} {\bibfield
  {journal} {\bibinfo  {journal} {Phys. Rev. B}\ }\textbf {\bibinfo {volume}
  {63}},\ \bibinfo {pages} {045115} (\bibinfo {year} {2001})}\BibitemShut
  {NoStop}%
\bibitem [{\citenamefont {Takeya}\ \emph {et~al.}(2004)\citenamefont {Takeya},
  \citenamefont {Togano}, \citenamefont {Sung}, \citenamefont {Mochiku},\ and\
  \citenamefont {Hirata}}]{Takeya2004}%
  \BibitemOpen
  \bibfield  {author} {\bibinfo {author} {\bibfnamefont {H.}~\bibnamefont
  {Takeya}}, \bibinfo {author} {\bibfnamefont {K.}~\bibnamefont {Togano}},
  \bibinfo {author} {\bibfnamefont {Y.~S.}\ \bibnamefont {Sung}}, \bibinfo
  {author} {\bibfnamefont {T.}~\bibnamefont {Mochiku}},\ and\ \bibinfo {author}
  {\bibfnamefont {K.}~\bibnamefont {Hirata}},\ }\bibfield  {title} {\bibinfo
  {title} {Metastable superconductivity in niobium diborides},\ }\href
  {https://doi.org/https://doi.org/10.1016/j.physc.2004.02.046} {\bibfield
  {journal} {\bibinfo  {journal} {Physica C: Superconductivity}\ }\textbf
  {\bibinfo {volume} {408-410}},\ \bibinfo {pages} {144} (\bibinfo {year}
  {2004})},\ \bibinfo {note} {proceedings of the International Conference on
  Materials and Mechanisms of Superconductivity. High Temperature
  Superconductors VII -- M2SRIO}\BibitemShut {NoStop}%
\end{thebibliography}%

\end{document}